\documentclass[notitlepage,english,aps,prx,tightenlines,floatfix,longbibliography,10pt,citeautoscript,twocolumn,colorlinks=true,urlcolor=blue,citecolor=blue, linkcolor=blue, superscriptaddress]{revtex4-2} 

\usepackage{graphicx}
\usepackage{amsmath,amssymb}
\usepackage{booktabs}
\usepackage{multirow}
\usepackage{array}
\usepackage{makecell}
\usepackage[T1]{fontenc}
\usepackage{lmodern}
\usepackage{cmap}
\usepackage{fix-cm}
\usepackage{enumitem}
\usepackage{orcidlink}
\usepackage{physics}   
\usepackage{xparse}
\usepackage{hyperref}

\usepackage{xr}
\newcommand{\cellBNR}[3]{\makecell{$(#1;#2)$\\$#3$}}

\NewDocumentCommand{\cellBNBNR}{ m m o m m o m }{
  \makecell{$(#1,#2\IfNoValueTF{#3}{}{,#3}) \approx (#4, #5\IfNoValueTF{#6}{}{, #6})$\\$#7$}}

\newcommand{\cellkBMR}[4]{\makecell{$(#1;#2,#3)$\\$#4$}}
\newcommand{\cellkBkBMR}[6]{\makecell{$(#1;#2) \sim (#3;#4,#5)$\\$#6$}}
\newcommand{\cellkBkBMRe}[6]{\makecell{$(#1;#2) = (#3;#4,#5)$\\$#6$}}
 
\newcommand{\cellR}[1]{\makecell{---\\$#1$}}

\newcommand{\cellText}[1]{\makecell{#1\\\ }}

\setcellgapes{1pt}\makegapedcells
\newcommand{\ham}{\mathcal{\hat H}}

\usepackage{fp}
\newcount\boxheight
\newcount\boxwidth
\newcommand\testaspect[1]{%
  \setbox0=\hbox{#1}%
  \boxheight=\ht0\relax%
  \boxwidth=\wd0\relax%
  \FPdiv\theaspect{\the\boxwidth}{\the\boxheight}%
  \copy0%
}

\begin{document}
\title{Parity-dependent double degeneracy and spectral statistics in the projected dice lattice}

\author{Koushik Swaminathan\orcidlink{0000-0003-4932-9977}}

\author{Anouar Moustaj\orcidlink{0000-0002-9844-2987}}

\author{Jose L. Lado\orcidlink{0000-0002-9916-1589}}

\author{Sebastiano Peotta\orcidlink{0000-0002-9947-1261}}
\email{sebastiano.peotta.work@protonmail.com}	
\affiliation{Department of Applied Physics, Aalto University, FI-00076 Aalto, Finland}

\begin{abstract}
We investigate the spectral statistics of an interacting fermionic system derived by projecting the Hubbard interaction onto the two lowest‑energy, degenerate flat bands of the dice lattice subjected to a $\pi$‑flux. Surprisingly, the distributions of level spacings and gap ratios correspond to distinct Gaussian ensembles, depending on the parity of the particle number. For an even number of particles, the spectra conform to the Gaussian Orthogonal Ensemble, as expected for a time‑reversal‑symmetric Hamiltonian. In stark contrast, the odd‑parity sector exhibits exact double degeneracy of all eigenstates even after resolving all known symmetries, and the Gaussian Unitary Ensemble accurately describes the spacing distribution between these doublets. The simultaneous emergence of two different random‑matrix ensembles within a single physical system constitutes an unprecedented finding, opening new avenues for both random matrix theory and flat‑band physics.
\end{abstract}

\maketitle

Flat electronic bands are a flexible building block for collective states of matter. In particular, in moir\'e materials, a slight twist between atomic layers produces nearly dispersionless bands and dramatically amplifies interaction effects~\cite{Cao_2018_Nature_Unconventionalsuperconductivitymagicangle, Balents_2020_Nat.Phys._Superconductivitystrongcorrelations, Andrei_2021_NatRevMater_marvelsmoirematerials}. These flat‑band platforms host a variety of correlated phases, such as unconventional superconductivity~\cite{Cao_2018_Nature_Unconventionalsuperconductivitymagicangle, Cao_2021_Nature_Paulilimitviolationreentrant, Park_2026_Science_Experimentalevidencenodal, Zhou_2021_Nature_Superconductivityrhombohedraltrilayer, Han_2025_Nature_Signatureschiralsuperconductivity} and fractional quantum Hall states~\cite{Wang_2015_Science_Evidencefractionalfractal, Lu_2024_Nature_Fractionalquantumanomalous, Kim_2025_NatCommun_Observation13,  Zhao_2025_ACSNano_ExploringFractionalQuantum}.

A common strategy to tackle the strongly interacting problem posed by lattice models with flat bands is to project the interaction term onto the flat‑band subspace, thereby significantly reducing the Hilbert‑space dimension~\cite{Huber_2010_Phys.Rev.B_Bosecondensationflat, Moller_2012_Phys.Rev.Lett._CorrelatedPhasesBosons, Derzhko_2015_Int.J.Mod.Phys.B_Stronglycorrelatedflatband,Tovmasyan_2016_Phys.Rev.B_Effectivetheoryemergent, Leykam_2018_AdvancesinPhysics:X_Artificialflatband}. The trade‑off is that the resulting projected Hamiltonian consists of a large number of nonlocal terms, which hampers both analytical treatment and large‑scale numerics~\cite{Leykam_2018_AdvancesinPhysics:X_Artificialflatband}. Moreover, purely quartic Hamiltonians obtained via flat‑band projection often exhibit emergent symmetries, quantum scars, and Hilbert‑space fragmentation~\cite{Tovmasyan_2018_Phys.Rev.B_Preformedpairsflat, Kuno_2020_NewJ.Phys._Flatbandmanybodylocalization, Nicolau_2023_Phys.Rev.B_Flatbandinduced, Swaminathan_2023_Phys.Rev.Res._Signaturesmanybodylocalization}, with profound effects on transport and many-body dynamics.

\begin{figure}[tb]
    \centering

    \includegraphics[width=\columnwidth]{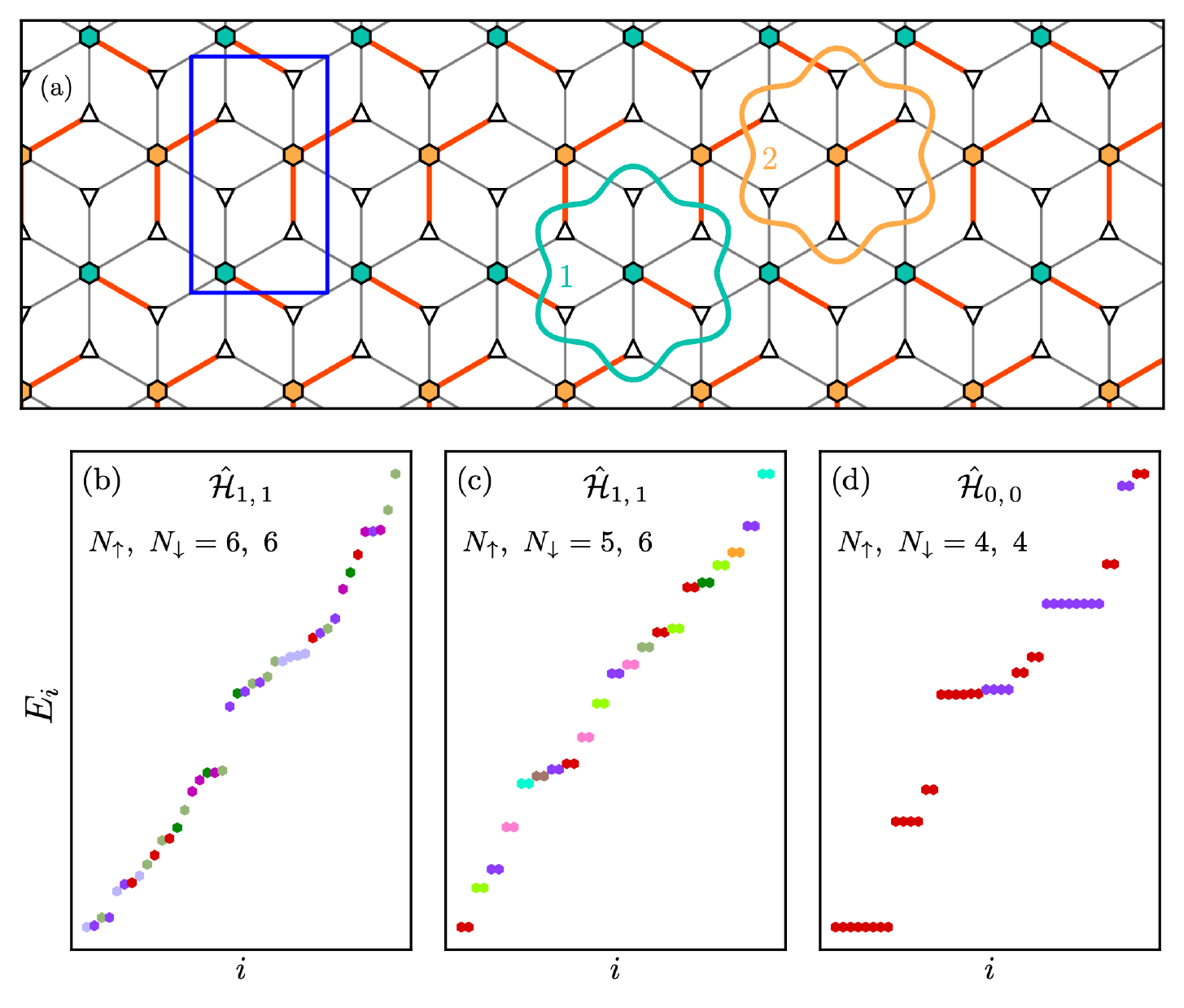}
    
    \caption{(a) Schematic of the dice lattice. The box is the magnetic unit cell comprising six lattice sites, while the thin black (thick red) bonds represent positive (negative) hopping amplitudes $t > 0$ ($-t <0$). Wannier functions $w_{n\mathbf{l}}$ ($n = 1, 2$) of the two lowest flat bands are also shown. 
    (b)-(d): Interior eigenvalues of $\ham_{\lambda_2, \lambda_3}$~\eqref{eq:H_main} for system size $N_x, N_y =3, 2$, momentum $\mathbf{k} = 0$, and particle numbers $N_\uparrow$, $N_\downarrow$ shown in each panel. Different colors indicate different pairs $(S, B)$ of total spin and pseudospin quantum numbers. For even particle numbers (b), the eigenvalues $E_i$ are nondegenerate. For odd particle numbers (c), the spectrum consists of degenerate doublets. When $\lambda_2=\lambda_3=0$ (d), highly degenerate multiplets appear due to known LIOMs~\cite{Swaminathan_2023_Phys.Rev.Res._Signaturesmanybodylocalization, supplementary}.}
    \label{fig:lat+spec}
\end{figure}

The dice lattice~\cite{Vidal_1998_Phys.Rev.Lett._AharonovBohmCagesTwoDimensional, Vidal_2001_Phys.Rev.B_DisorderinteractionsAharonovBohm, Moller_2012_Phys.Rev.Lett._CorrelatedPhasesBosons} (Fig.~\ref{fig:lat+spec}(a)) offers a highly tractable platform for investigating these effects. Its flat bands support compactly localized Wannier functions, which allow the projected Hamiltonian to be written as a finite collection of short‑range terms. This convenient structure has enabled detailed studies of the lattice’s many‑body physics, including prior reports of quasiparticle localization that may arise from emergent local integrals of motion (LIOMs)~\cite{Swaminathan_2023_Phys.Rev.Res._Signaturesmanybodylocalization}. Here, we examine the spectral statistics, specifically, the level spacing and gap ratios distributions, to resolve the open question of whether LIOMs are present in this system.

In generic nonintegrable systems, level spacing and gap ratio distributions follow the Gaussian orthogonal, unitary or symplectic ensembles (GOE/GUE/GSE) of random matrix theory (RMT)~\cite{Guhr_1998_PhysicsReports_Randommatrixtheoriesquantum, Haake_2010__QuantumSignaturesChaos, Mehta_2004__RandomMatrices}, while systematic deviations signal hidden symmetries, integrability, or fragmentation of the Hilbert space~\cite{Sala_2020_Phys.Rev.X_ErgodicityBreakingArising, Moudgalya_2022_Phys.Rev.X_HilbertSpaceFragmentation, Moudgalya_2022_Rep.Prog.Phys._Quantummanybodyscars}. In particular, higher-order ($k$-nearest-neighbor, $k$NN) spectral statistics have recently emerged as sensitive probes of multi-block spectra and hidden symmetries~\cite{Tekur_2018_Phys.Rev.B_Higherorderspacingratios, Rao_2020_Phys.Rev.B_Higherorderlevelspacings, He_2026_J.Stat.Mech._Statisticalsignaturesintegrable}.

In this work, we analyze the spectral statistics of the Hamiltonian obtained by restricting a Hubbard interaction to the two lowest degenerate flat bands of the dice lattice threaded by a $\pi$-flux. Exact diagonalization in symmetry-resolved sectors reveals a surprising parity dichotomy. For even total particle number, the spectrum is nondegenerate (see Fig.~\ref{fig:lat+spec}(b)) and the spectral statistics at any order $k$ correspond to the superposition of $m$ independent matrix blocks in the GOE. The number of blocks $m$ changes with system size, suggesting the presence of an extensive number of LIOMs, which would explain the nonergodic dynamics observed previously~\cite{Swaminathan_2023_Phys.Rev.Res._Signaturesmanybodylocalization}.
Instead, for odd parity, the spectral statistics display entirely unexpected features. 
The full spectrum is doubly degenerate even after resolving all known symmetries, such as spin rotation (see Fig.~\ref{fig:lat+spec}(c)), and for all accessible cluster sizes, the spacing distribution of the doublets ($k=2$ and $4$ order statistics) matches that of the GUE.

Our work highlights the usefulness of higher-order spectral statistics and, most importantly, reveals an unprecedented phenomenon, the coexistence of different random-matrix ensembles (GOE and GUE) in the same physical system, which is a minimal and paradigmatic lattice model with flat bands. More surprisingly, the switch from one ensemble to the other occurs simply by removing or adding a single particle. This suggests the existence of a new class of Hamiltonians that defy the expectations from RMT.

\textit{Model and symmetries—}The Wannier functions $w_{n\mathbf{l}}$ of the two lowest flat bands of the dice lattice with $\pi$-flux can be taken to be compactly localized~\cite{Vidal_1998_Phys.Rev.Lett._AharonovBohmCagesTwoDimensional, Vidal_2001_Phys.Rev.B_DisorderinteractionsAharonovBohm, Tovmasyan_2018_Phys.Rev.B_Preformedpairsflat, Zhang_2020_Phys.Rev.B_Compactlocalizedstates,Moller_2012_Phys.Rev.Lett._CorrelatedPhasesBosons}, as seen in Fig.~\ref{fig:lat+spec}(a). Consequently, the projection yields an effective interaction with a finite number of nearest-neighbor terms acting within the flat-band subspace, which we parametrize as
\begin{equation}
    \label{eq:H_main}
    \ham_{\lambda_2,\lambda_3}
      = \ham_{\mathrm{tri.}}
      + \lambda_2\,\ham_{\mathrm{kag.}}
      + \lambda_3\,\ham_{\mathrm{tri.-kag.}}.
\end{equation}
The first term $\ham_{\rm tri.}$ describes the motion of on-site pairs and the exchange interaction between spins, $\ham_{\rm kag.}$ controls the dynamics of pairs delocalized on two neighboring Wannier functions (bond singlets), which move in an effective kagome lattice, and $\ham_{\rm tri-kag.}$ is a term converting on-site into bond singlets and viceversa. The projected Hamiltonian of the dice lattice is realized for $\lambda_2=\lambda_3=1$. The explicit forms of all terms are given in Ref.~\cite{Swaminathan_2023_Phys.Rev.Res._Signaturesmanybodylocalization} and in the Supplemental Material~\cite{supplementary}.

The Hamiltonian Eq.~\eqref{eq:H_main} possesses many symmetries that should be resolved when analyzing its spectral statistics. Under periodic boundary conditions, $\ham_{\lambda_2, \lambda_3}$ commutes with the translation operators of a rectangular Bravais lattice. Thus, the eigenstates are labeled by the crystal momentum $\mathbf k=(k_x,k_y)$, with $0\leq k_\mu < N_\mu$ on a system composed of $N_x\times N_y$ unit cells. In the presence of the $\pi$-flux, point-group symmetries such as reflections and two-fold rotations are dressed by gauge transformations. Therefore, their respective quantum numbers cannot be resolved simultaneously with momentum $\mathbf k$~\cite{supplementary}.

The Hamiltonian possesses also full $\mathrm{SU(2)}$ spin symmetry since it commutes with the spin ladder operator $\hat{S}^+ = \sum_{n\mathbf{l}}\hat{d}^\dagger_{n\mathbf{l}\uparrow}\hat{d}_{n\mathbf{l}\downarrow}$ and the total spin operator $\hat{S}^2 = (S^z)^2  + \frac{1}{2}(\hat{S}^+\hat{S}^- + \hat{S}^-\hat{S}^+)$, where $\hat{d}_{n\mathbf{l}\sigma}$ is the fermionic annihilation operator corresponding to Wannier function $w_{n\mathbf{l}}$ and $\sigma = \uparrow,\,\downarrow$. When $N_\uparrow=N_\downarrow$, there is an additional spin-flip symmetry, $S^z\to -S^z$. Even (odd) states under spin flip are labeled by $f=+(-)$. Moreover, for $\lambda = \lambda_2=\lambda_3$ one has the relation $[\ham_{\lambda,\lambda}, \hat{B}^+] = -E_{\rm p}\hat{B}^+$ (called a spectrum generating algebra~\cite{Barut_1965_Phys.Rev._DynamicalGroupsMass, Buca_2019_NatCommun_Nonstationarycoherentquantum, Moudgalya_2022_Rep.Prog.Phys._Quantummanybodyscars}), where the operator $\hat{B}^+ = \sum_{n\mathbf{l}}\hat{d}^\dagger_{n\mathbf{l}\uparrow}\hat{d}^\dagger_{n\mathbf{l}\downarrow}$ creates an on-site pair in a zero momentum state. Even for nonzero binding energy $E_{\rm p}$, this implies that the eigenstates can be labeled by the eigenvalues of the form $B(B+1)$ of the total \textit{pseudospin} operator $\hat{B}^2 = (\hat{B}^z)^2 +  \frac{1}{2}(\hat{B}^+\hat{B}^- + \hat{B}^- \hat{B}^+)$~\cite{Tovmasyan_2016_Phys.Rev.B_Effectivetheoryemergent}. Numerically, the basis of a sector with given particle numbers  $N_\uparrow$ and $N_\downarrow$, momentum $\mathbf{k}$, and (if applicable) spin-flip parity $f$ is constructed using QuSpin~\cite{Weinberg_2017_SciPostPhys._QuSpinPythonpackage, Weinberg_2019_SciPostPhys._QuSpinPythonpackage},
and the spin $S$ and pseudospin $B$ quantum numbers are resolved by diagonalizing $\ham_{\lambda_2,\lambda_3}+\alpha \hat S^2+\beta \hat B^2$ with large $\alpha, \beta$~\cite{Poilblanc_1993_EPL_PoissonvsGOE,Teeriaho_2025_Phys.Rev.Res._Coexistenceergodicnonergodic}.

The term $\ham_{\mathrm{tri.}}$ in~\eqref{eq:H_main} commutes with an extensive number of LIOMs, namely the total spin operators $\hat{S}^2_{n\mathbf{l}}$ of each Wannier function~\cite{Swaminathan_2023_Phys.Rev.Res._Signaturesmanybodylocalization}. Whether these LIOMs are either broken or deformed for nonzero $\lambda_2$ and $\lambda_3$ is the open question that motivates this work. Finally, it is important to note that the Hamiltonian is time-reversal symmetric since all hopping matrix elements are real.

\textit{Spectral statistics—}For an ordered spectrum $\{E_n\}$, we define the $k$NN ($k$-order) level spacings
\begin{equation}
\tilde{s}_n^{\,k}=E_{n+k}-E_n , \qquad k=1,2,3,4.
\end{equation}
To remove the smooth energy dependence of the density of states, these spacings are unfolded and normalized according to $s_n^{\,k}={\tilde{s}_n^{\,k}}/{\langle \tilde{s}^{\,k}\rangle_{\rm local}}$,
such that $\langle s^{\,k}\rangle_{\rm local}\simeq1$
\cite{Santos_2004_J.Phys.A:Math.Gen._IntegrabilitydisorderedHeisenberg,
Teeriaho_2025_Phys.Rev.Res._Coexistenceergodicnonergodic,supplementary}. The resulting distributions $P^k(s)$ are then compared to the universal predictions of RMT. It is natural to consider the second and higher-order level spacings due to the degeneracy observed for odd parity. The distribution of $k$NN level spacings is also useful to identify independent blocks~\cite{Tekur_2018_Phys.Rev.B_Higherorderspacingratios,
Rao_2020_Phys.Rev.B_Higherorderlevelspacings}, as observed for even parity.
Complementary information is obtained from the distribution $P^k(r)$ of non-overlapping $k$-order gap ratios,
\begin{equation}\label{eq:gapratio}
r_n^{\,k}=
\frac{\min(\tilde{s}_n^{\,k},\tilde{s}_{n+k}^{\,k})}{\max(\tilde{s}_n^{\,k},\tilde{s}_{n+k}^{\,k})},
\end{equation}
which do not require unfolding
\cite{Oganesyan_2007_Phys.Rev.B_Localizationinteractingfermions,
Atas_2013_Phys.Rev.Lett._DistributionRatioConsecutive,
Tekur_2018_Phys.Rev.B_Higherorderspacingratios,Rao_2020_Phys.Rev.B_Higherorderlevelspacings}.

The spacing distributions are fitted to the function 
\begin{equation} \label{eq:kNN_wigner}
P^k(s,\beta)\approx C_{\alpha}\, s^{\alpha}
\exp[-A_{\alpha}s^2],
\end{equation}
with $\alpha$ as the only free parameter since the coefficients $A_\alpha$ and $C_\alpha$  are fixed by the conditions
$\int P^k(s)\dd{s} = 1$ and $\int sP^k(s) \dd{s} = \langle s\rangle = 1$. Within the simplest approximation of the Wigner surmise generalized to arbitrary order $k$, the exponent $\alpha$ is given by \cite{Tekur_2018_Phys.Rev.B_Higherorderspacingratios,
Rao_2020_Phys.Rev.B_Higherorderlevelspacings}
\begin{equation}\label{eq:alphabk}
\alpha(\beta,k)=\frac{k(k+1)}{2}\beta+k-1,
\end{equation}
where the Dyson index takes values $\beta = 1,2,4$ for the  GOE, GUE, GSE respectively. 
The effective index $\beta^*$ and order $k^*$, which best describe the numerical results, are obtained by comparing the fitted value $\alpha_{\rm fit}$ to an improved version of Eq.~\eqref{eq:alphabk}~\cite{Shir_2025_J.Phys.A:Math.Theor._Surmiserandommatrices, supplementary}. The analogue of Eqs.~\eqref{eq:kNN_wigner} and \eqref{eq:alphabk} for non-overlapping gap ratios is given in Refs.~\cite{Atas_2013_Phys.Rev.Lett._DistributionRatioConsecutive, Atas_2013_J.Phys.A:Math.Theor._Jointprobabilitydensities, Tekur_2018_Phys.Rev.B_Higherorderspacingratios}.

A complication arises when the spectrum is the superposition of $m$ independent subsequences, as can occur due to unresolved symmetries or Hilbert-space fragmentation. Levels belonging to different subsequences do not repel, causing 1NN statistics to appear closer to the Poisson distribution even when each subsequence is well described by a Gaussian ensemble~\cite{Bleher_2001__Randommatrixmodels, Anderson_2009__IntroductionRandomMatrices, Deift_2014__Randommatrixtheory, Giraud_2022_Phys.Rev.X_ProbingSymmetriesQuantum}. 
To determine $m$, in addition to the fitting procedure described above, we also compare against the distribution $P^k(\xi,\beta,m)$, which denotes the $k$th-order spectral statistic for a superposition of $m$ independent blocks with Dyson index $\beta$. Here, $\xi=s$ corresponds to the $k$NN spacing distribution, while $\xi=r$ corresponds to the non-overlapping $k$th-order gap-ratio distribution. We write $P^k(\xi,\beta,1)=P^k(\xi,\beta)$ for a single block. The distributions $P^k(\xi,\beta,m)$ are generated by Monte Carlo sampling, and the closest matching one as a function of $m$ is chosen by minimizing the Kolmogorov--Smirnov distance~\cite{Massey_1951_JournaloftheAmericanStatisticalAssociation_KolmogorovSmirnovTestGoodness, Stephens_1974_JournaloftheAmericanStatisticalAssociation_EDFStatisticsGoodness} and the mean-square deviation~\cite{supplementary}.

An additional subtlety is that distinct ensembles at different orders can yield identical spectral statistics. A prominent example is the equivalence between the 2NN statistics of two superposed GOE blocks and those of a single GUE one: $P^{2k}(\xi,1,2)=P^k(\xi,2)$~\cite{Bleher_2001__Randommatrixmodels, Tekur_2018_Phys.Rev.B_Higherorderspacingratios}. Other similar inter-relationships between multi-block distributions~\cite{Bleher_2001__Randommatrixmodels, Forrester_2004_Probab.TheoryRelat.Fields_Correlationssuperpositionsdecimations, Forrester_2009_Commun.Math.Phys._RandomMatrixDecimation, Tekur_2020_Phys.Rev.Res._Symmetrydeductionspectral, Bhosale_2021_Phys.Rev.B_Superpositionhigherorderspacing, He_2026_J.Stat.Mech._Statisticalsignaturesintegrable} are encountered below.

\textit{Results—}As shown in Fig.~\ref{fig:lat+spec}(b), for even total particle number $N = N_\uparrow + N_\downarrow$, the eigenvalues are nondegenerate. In contrast, in the odd-$N$ sector the spectrum is composed of exact doublets $(E_{2l},E_{2l+1}=E_{2l})$ [Fig.~\ref{fig:lat+spec}(c)]. This pairing occurs throughout the $(\lambda_2,\lambda_3)$ plane~\cite{supplementary} and survives after resolving all known mutually commuting conserved quantities $(N_\uparrow,N_\downarrow,k_x,k_y, f,S,B)$. It is interesting to note that local observables such as spin and particle densities are related within each degenerate doublet by a flip $1\leftrightarrow 2$ of the band (Wannier) index $n$~\cite{supplementary}. Since $S^z$ is fixed, these degeneracies are not due to the usual spin-$\tfrac12$ time-reversal symmetry. In the limit $\lambda_2=\lambda_3=0$, the spectrum  is highly degenerate [Fig.~\ref{fig:lat+spec}(d)], due to the LIOMs of $\ham_{0,0}$~\cite{Tovmasyan_2018_Phys.Rev.B_Preformedpairsflat, Swaminathan_2023_Phys.Rev.Res._Signaturesmanybodylocalization}.

\begin{figure*}
  \centering
  \includegraphics[width=\textwidth]{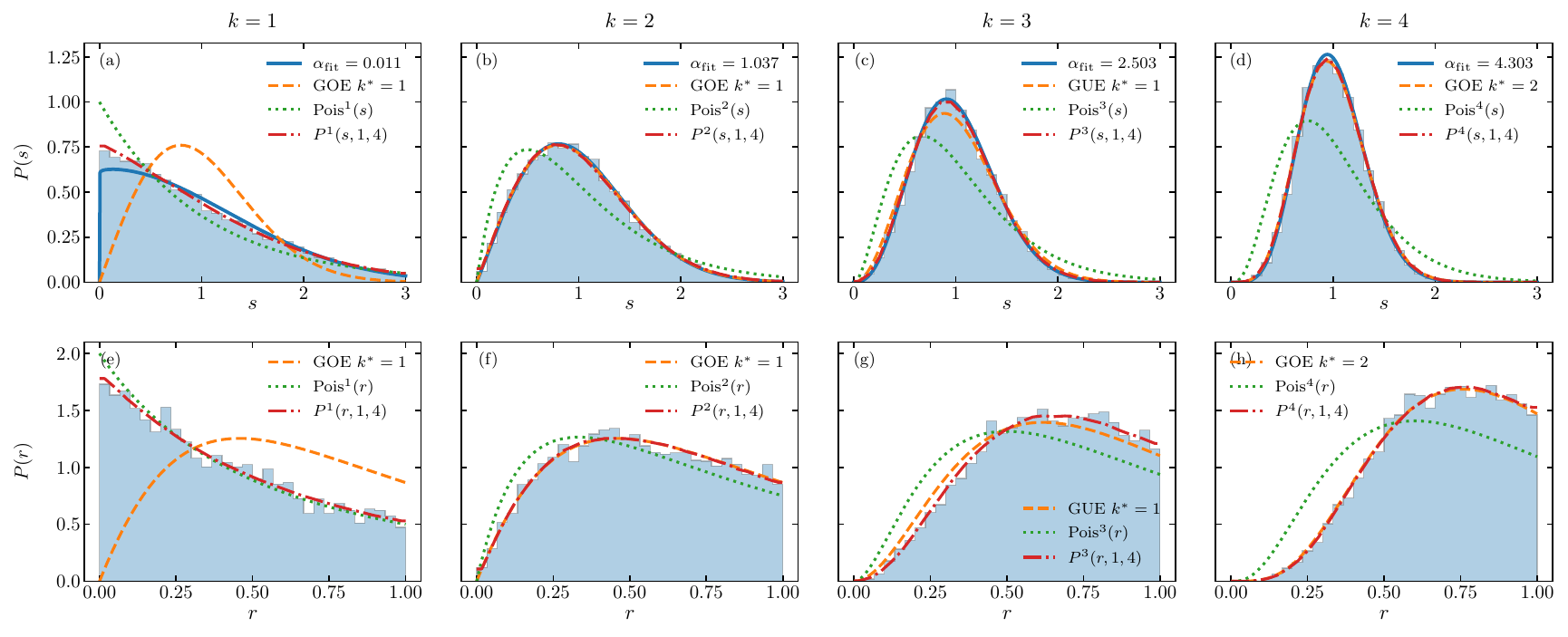}
  \caption{Level spacing (top) and gap ratio (bottom) statistics of $\ham_{1,1}$ for even-$N$, size $N_x, N_y=3, 2$, and quantum numbers $(N_\uparrow, N_\downarrow,k_x,k_y,f, S, B) = (6,6,0,0,+,0,1)$ (Hilbert-space dimension $\mathcal N=16529$). In the top row, the solid line shows the fit to \eqref{eq:kNN_wigner}. In both rows, the dashed curve is the closest matching Wigner surmise with effective order $k^*$ (ensemble $\beta^*$). The dotted curve shows the  $k$NN Poisson distribution, and the dash-dot curve is the closest $m$-block GOE distribution for each $k$. Both level-spacing and gap-ratio distributions are consistent with the superposition of four GOE spectra for any $k$. Note that $P(\xi, 1) = P^2(\xi, 1, 4)$ as seen in (b) and (f), while $P^2(\xi, 1) = P^4(\xi, 1, 4)$ from (d) and (h).}
  \label{fig:kNNeven}
\end{figure*}  

\begin{figure*}
  \centering
  \includegraphics[width=\textwidth]{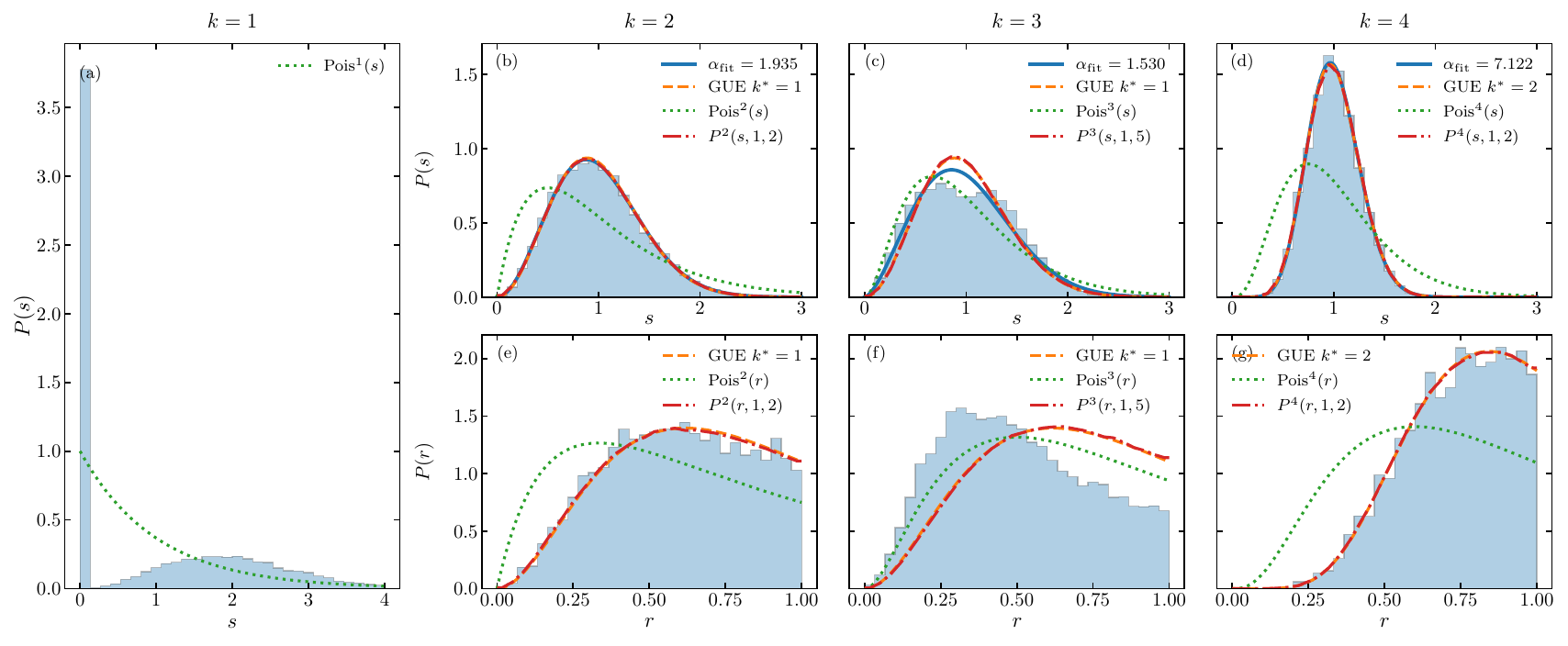}
  \caption{Same as Fig.~\ref{fig:kNNeven}, but for odd-$N$ and quantum numbers $(N_\uparrow,N_\downarrow,k_x,k_y,S,B) = (5,6,0,0,1/2,1/2)$ (dimension $\mathcal N=25176$). For $k = 1$, the double degeneracy leads to a pronounced peak at $s=0$ in the level-spacing distribution and identically zero gap ratio~\eqref{eq:gapratio} (not depicted). The $k=2,4$ statistics are very well described by the GUE with $k^* = 1, 2$, equivalent to the superposition of two GOE blocks due to the known relation $P^{2k^*}(\xi, 1, 2)=P^{k^*}(\xi, 2)$. However, the $k = 3$ case clearly deviates from the superposition of two GOE blocks. Note the relation $P(\xi, 2) = P^3(\xi, 1, 5)$ in (c) and (f).}
  \label{fig:kNNodd}
\end{figure*} 

A pronounced parity dependence appears in the spectral statistics. Figure~\ref{fig:kNNeven} shows the $k$NN level spacing and gap ratio distributions for an even-$N$ sector on an $N_x, N_y = 3, 2$ lattice. Both $P(s)$ and $P(r)$ are well described by $P^k(\xi, 1, 4)$, namely the superposition of four GOE blocks. For $k=1$, the distribution is close to Poissonian, especially seen in the gap ratio, while the difference becomes clearer at higher orders. This behavior is consistent with the presence of LIOMs, which may explain the observed signatures of quasiparticle localization~\cite{Swaminathan_2023_Phys.Rev.Res._Signaturesmanybodylocalization}. Fig.~\ref{fig:kNNeven} illustrates also the relations $P^1(\xi, 1) = P^2(\xi,1,4)$ and $P^2(\xi, 1) = P^1(\xi,4) = P^4(\xi,1,4)$ between different combinations $(k, \beta, m)$~\cite{Bleher_2001__Randommatrixmodels, Tekur_2020_Phys.Rev.Res._Symmetrydeductionspectral, Bhosale_2021_Phys.Rev.B_Superpositionhigherorderspacing}.

The odd-$N$ sector, shown in Fig.~\ref{fig:kNNodd} for the same lattice size $N_x, N_y=3,2$, displays qualitatively different behavior. The 1NN spacing distribution exhibits a pronounced peak at $s=0$ due to the complete double degeneracy, thus higher ($k>1$) order statistics are appropriate. Surprisingly, both 2NN and 4NN spacing and gap ratio distributions match the GUE at effective orders $k^*=1$ and $k^*=2$, respectively, which is equivalent to the superposition of two GOE blocks, since $P^{2k^*}(\xi,1,2) = P^{k^*}(\xi,2)$~\cite{Tekur_2018_Phys.Rev.B_Higherorderspacingratios, Bleher_2001__Randommatrixmodels}. By contrast, the 3NN spectral statistics does not resemble any $m$-block distribution, ruling out that the GUE behavior observed for $k= 2,4$ can be explained by the superposition of two GOE blocks.

\newcommand{\rowrulemain}{\cline{2-6}}
\squeezetable
\begin{table}[t]
\centering
\scriptsize
\renewcommand\arraystretch{1.15}

\begin{ruledtabular}
\begin{tabular}{ll*{4}{c}}
\multicolumn{1}{c}{} &
\textbf{k} &
$N_x, N_y=3,2$ &
$N_x, N_y=4,2$ &
$N_x, N_y=3,3$ &
\makecell[c]{sing. part. \\hopping}
\\
\hline

\multirow{4}{*}{\textbf{Even}}
& \textbf{1}    & \cellkBMR{1}{1}{4}{0.389} & \cellkBMR{1}{1}{6}{0.388} & \cellkBMR{1}{1}{2}{0.419}  & \cellBNR{1}{1}{0.528} \\
\rowrulemain
& \textbf{2}    & \cellkBkBMRe{1}{1}{2}{1}{4}{0.529} & \cellkBMR{2}{1}{6}{0.508} & \cellkBkBMRe{1}{2}{2}{1}{2}{0.599}  & \cellBNR{2}{1}{0.675} \\
\rowrulemain
& \textbf{3}    & \cellkBMR{3}{1}{4}{0.622} & \cellkBkBMR{1}{2}{3}{1}{6}{0.580} & \cellkBkBMRe{2}{1}{3}{1}{2}{0.680} & \cellBNR{3}{1}{0.746} \\
\rowrulemain
& \textbf{4}    & \cellkBkBMRe{2}{1}{4}{1}{4}{0.676} & \cellkBMR{4}{1}{6}{0.633} & \cellkBkBMRe{2}{2}{4}{1}{2}{0.731} & \cellBNR{4}{1}{0.789} \\

\cline{1-6}

\multirow{4}{*}{\textbf{Odd}}
& \textbf{1}    & \cellText{DD} & \cellText{DD} & \cellText{DD} & \cellBNR{1}{1}{0.530} \\
\rowrulemain
& \textbf{2}    & \cellkBkBMRe{1}{2}{2}{1}{2}{0.596} & \cellkBkBMRe{1}{2}{2}{1}{2}{0.598} & \cellkBkBMRe{1}{2}{2}{1}{2}{0.592} & \cellBNR{2}{1}{0.672} \\
\rowrulemain
& \textbf{3}    & \cellR{0.507} & \cellR{0.508} & \cellR{0.508} & \cellBNR{3}{1}{0.744} \\
\rowrulemain
& \textbf{4}    & \cellkBkBMRe{2}{2}{4}{1}{2}{0.731} & \cellkBkBMRe{2}{2}{4}{1}{2}{0.735} & \cellkBkBMRe{2}{2}{4}{1}{2}{0.732} & \cellBNR{4}{1}{0.791} \\
\end{tabular}
\end{ruledtabular}

\caption{\label{tab:lss}
Level statistics of the projected dice lattice Hamiltonian, grouped by the parity of $N=N_\uparrow+N_\downarrow$. Columns indicate the cluster size $(N_x, N_y)$ (and the single-particle-hopping-augmented case for system size $3\times 2$, last column); rows correspond to different orders $k=1,2,3,4$. Each cell reports, on the top line, the best-matching RMT identification either as an effective single block $(k^*;\beta^*)$ or as the superposition of $m$  blocks $(k;\beta,m)$. The equality sign denotes relations between different multi-block distributions~\cite{Bhosale_2021_Phys.Rev.B_Superpositionhigherorderspacing, Tekur_2020_Phys.Rev.Res._Symmetrydeductionspectral}. The bottom line gives the $k$-order mean nonoverlapping gap ratio $\langle r^{\,k}\rangle$ computed in the largest accessible symmetry-resolved sector. DD denotes exact double degeneracy of all levels.}
\end{table} 

The phenomenology of the observed spectral statistics for different system sizes is summarized in Tab.~\ref{tab:lss}. In the even-$N$ sector, all clusters studied display level statistics consistent with multi-block GOE behavior. Crucially, the inferred block number $m$ varies with the size, consistent with the presence of an extensive number of LIOMs, possibly at the root of the nonergodic behavior observed previously~\cite{Swaminathan_2023_Phys.Rev.Res._Signaturesmanybodylocalization}. However, the number of blocks does not increase monotonically with system size, being $m = 6$ for the $4\times 2$ cluster and just $m=2$ for $3\times 3$~\cite{supplementary}. This may be a finite-size effect due to the small system sizes that are feasible with exact diagonalization. By contrast, both the complete double degeneracy and GUE behavior in the 2NN and 4NN statistics are present in the odd-$N$ sector for all system sizes. Again, the 3NN distributions and the double degeneracy are incompatible with the superposition of two GOE blocks, and an alternative explanation is required for the GUE behavior.

Finally, the last column of Table~\ref{tab:lss} illustrates how the addition of a quadratic hopping term~\cite{supplementary} leads to a rapid loss of all parity- and size-dependent effects, rapidly driving the spectral statistics towards a single-block GOE distribution expected for a generic time-reversal and rotationally symmetric Hamiltonian without unresolved symmetries. Indeed, the complete double degeneracy of the spectrum, the GUE correlations in the 2NN and 4NN distributions for odd parity, and the size-dependent block number $m$ for even parity are exceptional and interesting features of the present model.  

\textit{Discussion—}A key open problem is to identify the mechanism enforcing the complete double degeneracy of the spectrum for odd parity. This is crucial since it most likely underpins all other unusual phenomena observed in the spectral statistics. At present, we have uncovered at least two hints towards the solution. The first is that the states in a degenerate doublet are related by the exchange of the two nonequivalent Wannier functions shown in Fig.~\ref{fig:lat+spec}(a), in the same way as the states in a Kramers pair are related by a spin flip $S^z\to -S^z$~\cite{Sakurai_2020__ModernQuantumMechanics}.
This would be consistent with a hidden antiunitary symmetry involving the band degree of freedom that would explain the degeneracy through Kramers' theorem. 

The second piece of evidence is that the double degeneracy is present in the full dice lattice with a Hubbard interaction, even before the projection to the flat-band subspace. This is useful as it may be easier to pinpoint the antiunitary symmetry in the full model rather than in the projected one. This observation also suggests that it would be interesting to perform the spectral statistics analysis on the full model to see if the GUE spectral correlations and the block structure persist. Due to the large unit cell of the dice lattice with $\pi$-flux, it is not feasible to perform a full diagonalization of the Hamiltonian; one has to employ methods that can compute at least a few thousand eigenvalues in a large Hilbert space~\cite{ Pietracaprina_2018_SciPostPhys._Shiftinvertdiagonalizationlarge, Sierant_2020_Phys.Rev.Lett._PolynomiallyFilteredExact, Guan_2021_SciPostPhys._DualapplicationsChebyshev}.

Alternatively, an intriguing possibility is that the degeneracy may be of topological origin. A relevant example is the Hamiltonian associated with the toric code~\cite{Kitaev_2003_AnnalsofPhysics_Faulttolerantquantumcomputation}, since all of its eigenvalues (not just the ground state) are $4^g$-fold degenerate on a surface of genus $g$. This would imply that this minimal interacting flat-band model features a parity-dependent topological order.

We also note a useful comparison with SYK-type Majorana models, where spectral statistics depend on $N_\chi \bmod 8$, with $N_\chi$ the number of Majorana modes defining the model. This dependence follows from the Bott periodicity of  Clifford algebras~\cite{Garcia-Garcia_2016_Phys.Rev.D_Spectralthermodynamicproperties, You_2017_Phys.Rev.B_SachdevYeKitaevmodelthermalization}. This should be distinguished from a dependence on particle number: in the SYK model, $N_\chi$ counts degrees of freedom of the Hamiltonian, and the model does not conserve particle number. The effect studied here is different: in the projected dice lattice, the number of single-particle orbitals is fixed, and the ensemble changes with the parity of the particle number. Thus, the SYK $N_\chi \bmod 8$ structure is a mode-number effect, whereas the present work concerns different particle-number sectors of a fixed lattice Hamiltonian. The phenomenology also differs: in the SYK model, changing $N_\chi \bmod 8$ cycles the ensemble among the standard Wigner--Dyson classes, whereas in the dice lattice we find multi-block GOE statistics in even particle-number sectors and exact doublets with GUE-like inter-doublet statistics in odd sectors. The microscopic origin of this parity dependence therefore remains an open question tied to the structure of the projected Hamiltonian, rather than the SYK Bott-periodicity mechanism.

Another important open question is whether the observed spectral features are exclusive to the dice lattice or can be found in other lattice models with flat bands. Of particular interest are models realized experimentally, such as the kagome and Lieb lattices~\cite{Jo_2012_Phys.Rev.Lett._UltracoldAtomsTunable, Thomas_2017_Phys.Rev.Lett._MeanFieldScalingSuperfluid, Leung_2020_Phys.Rev.Lett._InteractionEnhancedGroupVelocity, Taie_2015_Sci.Adv._Coherentdrivingfreezing, Taie_2020_NatCommun_Spatialadiabaticpassage} or
bands of moir\'e materials. Note that realistic proposals for realizing the dice lattice with $\pi$-flux using ultracold gases in optical lattices have been put forward~\cite{Andrijauskas_2015_Phys.Rev.A_ThreelevelHaldanelikemodel, Moller_2018_NewJ.Phys._Syntheticgaugefields, Tassi_2024_Adv.Phys.Res._ImplementationCharacterizationDice}, prompting further investigations on how the spectral features uncovered in this work may be observed in ultracold quantum simulators. A promising observable to consider is the particle counting statistics, which in a recent experiment with trapped Fermi gases are found in remarkable agreement with RMT predictions~\cite{Dixmerias_2025__UniversalRandomMatrix}.

\textit{Acknowledgments—}We acknowledge support from the Research Council of Finland under Grants No. 330384, No. 336369, and No. 358150, the Finnish Ministry of Education and Culture through the Quantum Doctoral Education Pilot Program (QDOC VN/3137/2024-OKM-4), and the Finnish Quantum Flagship (project No. 358877). We acknowledge the computational resources provided by the Aalto Science-IT project.

\textit{Data availability—}The data that support the findings of this study were generated by numerical simulations. The source code used to perform the simulations for this study as well as the datasets for the results in Tab.~\ref{tab:lss} are publicly available~\cite{Swaminathan_2026__Paritydependentdoubledegeneracya}.

 $ $

\end{document}


\title{Supplemental Material for ``Parity-dependent double degeneracy and spectral statistics in the projected dice lattice''}

\maketitle

This supplemental section presents the explicit representation and symmetry properties of the projected dice lattice Hamiltonian studied in the main text, along with details on the unfolding procedure, spectral statistics analysis, and additional supporting results, in particular, the spectral statistics analysis for the different system sizes used to fill the entries of Tab.~I in the main text. We follow the notation of 
Ref.~\cite{Swaminathan_2023_Phys.Rev.Res._Signaturesmanybodylocalization}.

\section{Projected dice lattice Hamiltonian}
\label{sec:model_app}

\begin{figure}[tb]
  \centering
    \includegraphics[width=0.32\columnwidth]{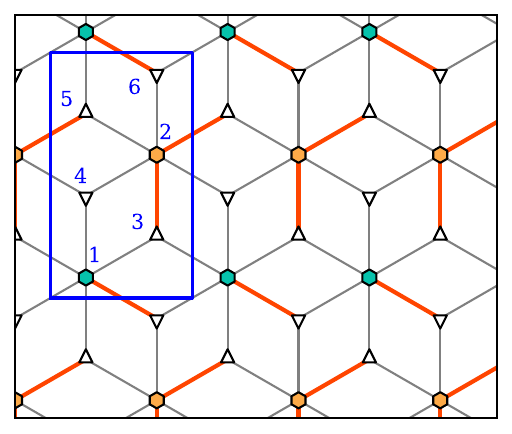}
  \includegraphics[width=0.64\columnwidth]{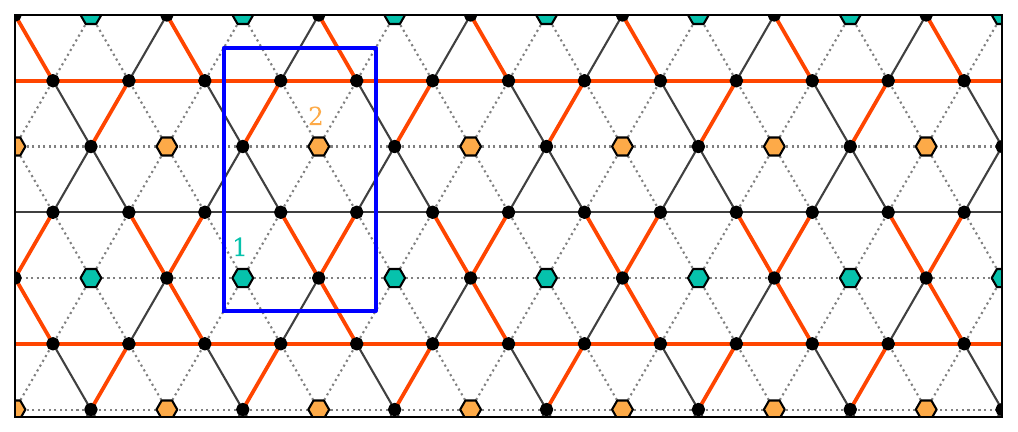}
  \caption{\label{fig:full_lattice} (left) Schematic of the dice lattice. The magnetic unit cell is highlighted in the rectangular box with orbital indices $\alpha = 1,2,3,4,5,6$. (right) Graphical representation of the term $\mathcal{\hat{H}}_{\rm tri.}+\mathcal{\hat{H}}_{\rm kag.}$ in the projected Hamiltonian $\hat{\mathcal H}_{\lambda_2,\lambda_3}$~\eqref{eq:H_lambdas_app}. The Wannier functions $w_{n = 1,\vb{l}}$ and $w_{n = 2,\vb{l}}$ are represented by the turquoise and orange hexagons, respectively. One unit cell (shown as the blue rectangle) contains one turquoise site and one orange site, corresponding to the two nonequivalent Wannier functions. The dotted lines connecting turquoise and orange sites represent the nearest-neighbor interaction and hopping terms of on-site pairs and on-site spins in $\mathcal{\hat{H}}_{\rm tri.}$~\eqref{eq:Htri_app}. In particular, the on-site pairs hop between nearest-neighbor sites on the triangular lattice composed of turquoise and orange sites. The black sites lie at the centers of the bonds connecting the turquoise and orange sites and form a kagome lattice. The operator $\hat{B}^+_{\langle n_1\vb{l}_1, n_2\vb{l}_2\rangle}$ creates a singlet on the bond $\langle n_1\vb{l}_1, n_2 \vb{l}_2\rangle$, thus the bond singlets live on the kagome lattice formed by the black dots. The black and red bonds represent terms of the form $\hat{B}^+_{\langle n_1\vb{l}_1,n_2\vb{l}_2\rangle}\hat{B}_{\langle n_1\vb{l}_1,n_3\vb{l}_3\rangle}^-$ in the bond-singlet hopping Hamiltonian $\mathcal{\hat{H}}_{\rm kag.}$~\eqref{eq:Hkag_app}. The sign of the hopping amplitude of the bonds connecting the black sites is given by $-s(n_1\vb{l}_1|n_2\vb{l}_2,n_3\vb{l}_3)$ and is equal to $-1$ for the black bonds and to $+1$ for the red bonds. The terms of the form $\hat{B}_{n_1\vb{l}_1}^+\hat{B}_{\langle n_2\vb{l}_2,n_3\vb{l}_3\rangle}^-$ in $\hat{\mathcal{H}}_{\rm tri.-kag.}$~\eqref{eq:Htrikag_app} are not represented in this figure for clarity. Refer to the appendix in Ref.~\cite{Swaminathan_2023_Phys.Rev.Res._Signaturesmanybodylocalization} instead.}
\end{figure}

We consider the dice lattice threaded by a $\pi$-flux per rhombus,
for which all bands are flat and two by two degenerate. The dice lattice single-particle Hamiltonian is represented graphically in the left panel of Fig.~\ref{fig:full_lattice}.
Projecting an attractive on-site Hubbard interaction onto the subspace corresponding to the two lowest flat bands yields an effective many-body Hamiltonian acting on a much smaller Hilbert space, thus enabling exact diagonalization of finite clusters with sufficiently large size. For practical
computations, it is convenient to work with an orthonormal Wannier basis of the
projected subspace and the corresponding canonical fermionic operators.
The flat bands admit compactly localized Wannier
functions $w_{n\vb l}$, where $n=1,2$ labels the two Wannier orbitals per unit
cell and the pair of integers $\vb l =(l_x, l_y)$ labels the unit cell. We denote by $\hat d_{n\vb l\sigma}$ the fermionic annihilation
operator for a fermion with spin $\sigma=\uparrow,\downarrow$ in Wannier state
$w_{n\vb l}$. Compact localization of $w_{n\vb l}$ is a key
simplifying feature of the dice lattice and underlies the short-range structure
of the projected Hamiltonian~\cite{Tovmasyan_2018_Phys.Rev.B_Preformedpairsflat, Zhang_2020_Phys.Rev.B_Compactlocalizedstates, Swaminathan_2023_Phys.Rev.Res._Signaturesmanybodylocalization}.

For each Wannier state, we define the spin-resolved and total  density operators
\begin{equation}
\hat\rho_{n\vb l\sigma}=\hat d^\dagger_{n\vb l\sigma}\hat d_{n\vb l\sigma},\qquad \hat\rho_{n\vb l}=\hat\rho_{n\vb l\uparrow}+\hat\rho_{n\vb l\downarrow},
\end{equation}
on-site spin operators
\begin{equation}
\hat S^+_{n\vb l}=\hat d^\dagger_{n\vb l\uparrow}\hat d_{n\vb l\downarrow},\quad
\hat S^-_{n\vb l}=(\hat S^+_{n\vb l})^\dagger,
\quad \hat S^x_{n\vb l}=\frac12(\hat S^+_{n\vb l}+\hat S^-_{n\vb l}), 
\quad \hat S^y_{n\vb l}=\frac{1}{2i}(\hat S^+_{n\vb l}-\hat S^-_{n\vb l}),
\quad \hat S^z_{n\vb l}=\frac12(\hat\rho_{n\vb l\uparrow}-\hat\rho_{n\vb l\downarrow}),
\end{equation}
and on-site pair (pseudospin) operators
\begin{equation}
\hat B^+_{n\vb l}=\hat d^\dagger_{n\vb l\uparrow}\hat d^\dagger_{n\vb l\downarrow}, \quad\hat B^-_{n\vb l}=(\hat B^+_{n\vb l})^\dagger,
\quad \hat B^x_{n\vb l}=\frac12(\hat B^+_{n\vb l}+\hat B^-_{n\vb l}), 
\quad \hat B^y_{n\vb l}=\frac{1}{2i}(\hat B^+_{n\vb l}-\hat B^-_{n\vb l}),
\quad \hat B^z_{n\vb l}=\frac12(\hat\rho_{n\vb l}-1).
\end{equation}
Each of the sets of operators $\hat{\vb S}_{n\vb l} = (\hat S^x_{n\vb l}, \,, \hat S^y_{n\vb l},\, \hat S^z_{n\vb l})^T$ and $\hat{\vb B}_{n\vb l}=(\hat B^x_{n\vb l}, \,, \hat B^y_{n\vb l},\, \hat B^z_{n\vb l})^T$ generates an
SU(2) algebra and they commute with each other $[\hat S^\alpha_{n\vb{l}},\hat B^\beta_{n'\vb{l}'}] = 0$.

Pairs of particles living on nearest-neighbor Wannier functions, called bond singlets, are created by the operator
\begin{equation}
\hat B^+_{\langle n_1\vb l_1,n_2\vb l_2\rangle}
=\hat d^\dagger_{n_1\vb l_1\uparrow}\hat d^\dagger_{n_2\vb l_2\downarrow}
-\hat d^\dagger_{n_1\vb l_1\downarrow}\hat d^\dagger_{n_2\vb l_2\uparrow},
\end{equation}
with $\hat B^-_{\langle n_1\vb l_1,n_2\vb l_2\rangle}$ its Hermitian conjugate. Here $\langle n_1\vb l_1,n_2\vb l_2\rangle$ denotes a pair of overlapping nearest-neighbor Wannier functions, namely Wannier functions whose supports overlap on exactly two sites of the dice lattice.

The Hamiltonian obtained by projecting on the subspace of the two lowest flat bands can be presented as the sum of three contributions
\begin{equation}
\label{eq:H_lambdas_app}
\hat{\mathcal H}_{\lambda_2,\lambda_3}
=\hat{\mathcal H}_{\mathrm{tri.}}
+\lambda_2\,\hat{\mathcal H}_{\mathrm{kag.}}
+\lambda_3\,\hat{\mathcal H}_{\mathrm{tri.-kag.}}.
\end{equation}
For convenience, the second and third terms are multiplied by the parameters $\lambda_2$ and $\lambda_3$. The Hamiltonian obtained by the projection on the flat-band subspace is given by the choice $\lambda_2=\lambda_3=1$.

The ``triangular'' term acts on spins and pseudospins on the triangular
lattice of Wannier centers:
\begin{equation}
\label{eq:Htri_app}
\begin{split}
\hat{\mathcal H}_{\mathrm{tri.}}
&= -A \sum_{n,\vb l}\hat B^+_{n\vb l}\hat B^-_{n\vb l}
-4\sum_{\langle n\vb l, n'\vb l'\rangle}
\bigg[
\Big(\hat B^z_{n\vb l}+\tfrac12\Big)\Big(\hat B^z_{n'\vb l'}+\tfrac12\Big)
+\frac12\Big(\hat B^+_{n\vb l}\hat B^-_{n'\vb l'}+\hat B^-_{n\vb l}\hat B^+_{n'\vb l'}\Big)
\bigg] \\
&\quad
+4\sum_{\langle n\vb l, n'\vb l'\rangle}
\bigg[
\hat S^z_{n\vb l}\hat S^z_{n'\vb l'}
+\frac12\Big(\hat S^+_{n\vb l}\hat S^-_{n'\vb l'}+\hat S^-_{n\vb l}\hat S^+_{n'\vb l'}\Big)
\bigg],
\end{split}
\end{equation}
where $A>0$ sets the on-site pair binding energy. The value of $A$  depends on the ratio between the Hubbard couplings on the three- and six-fold coordinated sites of the dice lattice (see left panel of Fig.~\ref{fig:full_lattice}) and does not have a qualitative effect on the results of this work, therefore it is always set to $A=10$.
 
The ``kagome'' term describes the hopping of bond singlets living on the
kagome lattice dual to the triangular bond network:
\begin{equation}
\label{eq:Hkag_app}
\hat{\mathcal H}_{\mathrm{kag.}}
=-\sum_{\langle 1,2,3\rangle}
\bigg[
s(1|2,3)\Big(\hat B^+_{\langle 1,2\rangle}\hat B^-_{\langle 1,3\rangle}
+\mathrm{H.c.}\Big)
+\text{cyclic perm. of }(1,2,3)
\bigg],
\end{equation}
where $\langle 1,2,3\rangle$ runs over unordered triplets of 
nearest-neighbor Wannier orbitals forming a triangle and the sign $s(1|2,3)=\pm1$ encodes the
overlap-induced sign structure controlling bond-singlet motion (see right panel of Fig.~\ref{fig:full_lattice}).

Finally, the ``triangular-kagome'' term converts on-site pairs into bond singlets and
vice versa:
\begin{equation}
\label{eq:Htrikag_app}
\hat{\mathcal H}_{\mathrm{tri.-kag.}}
=-\sum_{\langle 1,2,3\rangle}
\bigg[
s(1|2,3)\Big(\hat B^+_{1}\hat B^-_{\langle 2,3\rangle}
+\mathrm{H.c.}\Big)
+\text{cyclic perm. of }(1,2,3)
\bigg].
\end{equation}

\subsection{Symmetries and symmetry resolution used in exact diagonalization}
\label{sec:symmetry_app}

The projected Hamiltonian $\ham_{\lambda_2,\lambda_3}$ inherits a set of spatial and
internal symmetries whose resolution is necessary before analyzing the level spacing statistics (LSS).

\paragraph{Translations}
The lattice translation operators $\hat T_\mu$ ($\mu=1,2$) satisfy the following commutation relations
\begin{equation}
\hat T_\mu\,\hat d_{n\vb l\sigma}\,\hat T_\mu^\dagger
=\hat d_{n,\vb l+\vb e_\mu,\sigma},
\qquad [\ham_{\lambda_2,\lambda_3},\hat T_\mu]=0,
\end{equation}
where $\vb{e}_x = (1, 0)^T$ and $\vb{e}_y = (0, 1)^T$.
On a system comprising of $N_x\times N_y$ unit cells, their eigenvalues take the form  $\exp(2\pi i k_\mu/N_\mu)$, where $k_\mu$ are integers satisfying $0 \leq k_\mu \leq N_\mu$. The vector $\vb{k} = (k_x, k_y)$ is called here the crystal momentum, or simply momentum.

\paragraph{Reflections and twofold rotation}
As a consequence of the background $\pi$-flux, reflections are implemented as
gauge-dressed permutation operations $\hat R_\nu=\hat P_\nu \hat G_\nu$ where $\hat P_\nu$ permutes the sites/orbitals according to a mirror symmetry and $\hat G_\nu$ is a diagonal $\mathbb Z_2$ gauge transformation. We use the convention that the $\hat R_\nu$ reflection flips the unit-cell $\nu$-index.  
It is convenient to define the gauge-dressed reflections in the single-particle Hilbert space of the full dice lattice spanned by the single-particle basis $\ket{i_x,i_y,\alpha}$ (unit-cell coordinates
$(i_x,i_y)\in\mathbb Z^2$ and sublattice index $\alpha\in\{1,\dots,6\}$, see left panel of Fig.~\ref{fig:full_lattice}). Let the symbols $R_\nu$, $P_\nu$ and $G_\nu$ (without the ``hat'')  denote the single-particle operators corresponding to the Fock-space operators  $\hat R_\nu$, $\hat P_\nu$ and $\hat G_\nu$.
Then, the permutation operator $P_\nu$ is defined as
\begin{equation}
P_x \ket{i_x,i_y,\alpha}=
\begin{cases}
\ket{-i_x,\,i_y,\,\alpha}, & \alpha\in\{1,4,5\},\\
\ket{-i_x-1,\,i_y,\,\alpha}, & \alpha\in\{2,3,6\},
\end{cases}
\label{eq:Px_def_app}
\end{equation}
for the mirror symmetry flipping $i_x$, while for $i_y$ one has
\begin{equation}
P_y \ket{i_x,i_y,\alpha}=
\begin{cases}
\ket{i_x,\,-i_y,\,1}, & \alpha=1,\\
\ket{i_x,\,-i_y-1,\,2}, & \alpha=2,\\
\ket{i_x,\,-i_y-1,\,9-\alpha}, & \alpha\in\{3,4,5,6\},
\end{cases}
\end{equation}

The gauge transformations restore the $\pi$-flux hopping-sign pattern after the mirror symmetry has been applied and are defined by
\begin{equation}
G_x \ket{i_x,i_y,\alpha}=g_x(i_y,\alpha)\ket{i_x,i_y,\alpha},\qquad
g_x(i_y,\alpha)=
\begin{cases}
(-1)^{i_y}, & \alpha\in\{1,2,3,4,6\},\\
(-1)^{i_y+1}, & \alpha=5,
\end{cases}
\end{equation}
and
\begin{equation}
G_y \ket{i_x,i_y,\alpha}=g_y(i_x,\alpha)\ket{i_x,i_y,\alpha},\qquad
g_y(i_x,\alpha)=
\begin{cases}
(-1)^{i_x}, & \alpha\in\{1,2,4,5\},\\
(-1)^{i_x+1}, & \alpha\in\{3,6\}.
\end{cases}
\end{equation}
Note that $G_{x(y)}$ can be defined only if $N_{y(x)}$ is even. 
The commutation relations between the mirror symmetry operators and the translation operators are the following
\begin{equation}
\hat R_\nu \hat T_\nu \hat R_\nu^{-1}=\hat T_\nu^{-1},\qquad
\hat R_\nu \hat T_{\mu\neq\nu}\hat R_\nu^{-1}=-\hat T_\mu,
\qquad \mu,\nu=x,y.
\end{equation}
The minus sign on the right-hand side of the second relation is a consequence of the gauge transformation $\hat G_{\nu}$.
The product of the two reflections $\hat C_2=\hat R_x\hat R_y$ acts as a twofold rotation satisfying
\begin{equation}
\hat C_2 \hat T_\mu \hat C_2^{-1}=-\,\hat T_\mu^{-1},
\end{equation}
so that in momentum space $\hat C_2$ maps $\vb k\to-\vb k + (N_x/2, N_y/2)$ up to a reciprocal-lattice shift. As a result, $\hat C_2$
does not yield an independent quantum number in a generic momentum block,
except at momenta invariant under this mapping.

\paragraph{Spin $\mathrm{SU(2)}$ symmetry}
The model is invariant under global SU(2) spin rotations. Defining
$\hat S^\alpha=\sum_{n,\vb l}\hat S^\alpha_{n\vb l}$ and
$\hat{S}^{\,2}=\sum_{\alpha = x,y,z}(\hat S^\alpha)^2$, one has
\begin{equation}
[\ham_{\lambda_2,\lambda_3},\hat S^\alpha]=0,\qquad
[\ham_{\lambda_2,\lambda_3},\hat{S}^{\,2}]=0.
\end{equation}
In the numerics, we fix $S^z=N_\uparrow-N_\downarrow$ by working in a sector with a fixed number of particles $N_\uparrow$ and $N_\downarrow$. 
When $N_\uparrow = N_\downarrow$, there is an additional spin-flip symmetry $S^z \to -S^z$ that further splits the space into two sectors labelled by $ f \in \{+, -\}$.

\paragraph{Spectrum-generating algebra and pseudospin}
An important feature of $\ham_{\lambda_2,\lambda_3}$ for $\lambda_2 = \lambda_3 = \lambda$ is the
existence of the pair creation operator
\begin{equation}
\hat B^+=\sum_{n,\vb l}\hat B^+_{n\vb l}
=\sum_{n,\vb l}\hat d^\dagger_{n\vb l\uparrow}\hat d^\dagger_{n\vb l\downarrow},
\end{equation}
which acts as a ladder operator
\begin{equation}
\label{eq:bdagger_comm_app}
[\ham_{ \lambda, \lambda},\hat B^+]
=-E_{\mathrm p}\,\hat B^+,
\end{equation}
where $E_{\rm p}$ is the pair binding energy.
This property allows the construction of broad families of exact eigenstates generated
by repeated application of $\hat B^+$.
A consequence of~\eqref{eq:bdagger_comm_app} is that the total pseudospin operator
$\hat{B}^{\,2}=\sum_{\alpha = x,y,z}(\hat B^\alpha)^2$, with $\hat B^\alpha = \sum_{n,\vb{l}}\hat B_{n\vb{l}}^\alpha$, is also a conserved quantity.

\paragraph{Local integrals of motion of $\ham_{0,0}$}
The triangular Hamiltonian alone ($\hat{\mathcal H}_{\mathrm{tri.}} = \ham_{0,0}$) possesses an extensive set of strictly local conserved quantities. These are the total spin operators $\hat{{S}}_{n\vb{l}}^2 = \sum_{\alpha = x,y,z}  (\hat S_{n\vb{l}}^\alpha)^2$ of each Wannier function.
This means that the occupancy pattern of singly occupied Wannier orbitals is
conserved since the eigenvalue of $\hat{{S}}_{n\vb{l}}^2$ is $3/4$ if exactly one particle is present in the Wannier function labeled by $n\vb{l}$ and zero if it is empty or doubly occupied. These local integrals of motions (LIOMs) do not commute with $\ham _{\lambda_2,\lambda_3}$ if $\lambda_2$ or $\lambda_3$ is nonzero, that is, when bond-singlet hopping and conversion are present, but they motivate the search for deformed LIOMs organizing the spectrum away from $\lambda_2=\lambda_3=0$. Note that, due to the identity $\hat{{S}}_{n\vb{l}}^2 + \hat{{B}}_{n\vb{l}}^2 = \frac{3}{4}\hat{1}$, the operators $\hat{{B}}_{n\vb{l}}^2$ do not form a set of independent conserved quantities.

\paragraph{Time-reversal symmetry}
Since all Peierls phase factors in the single-particle dice lattice Hamiltonian are real, even after the projection, the model
is invariant under the standard antiunitary time-reversal  operator $\hat{\mathcal T}$ for spin-$\tfrac12$,
\begin{equation}
\hat{\mathcal T}\hat d_{n\vb l\uparrow}\hat{\mathcal T}^{-1}=\hat d_{n\vb l\downarrow},\qquad
\hat{\mathcal T}\hat d_{n\vb l\downarrow}\hat{\mathcal T}^{-1}=-\hat d_{n\vb l\uparrow},\qquad 
\hat{\mathcal T}^2=-1, \qquad \mathcal{\hat T}z = z^*\mathcal{\hat T}\qc z\in \mathbb{C}\,.
\end{equation}
Since $\hat{\mathcal T}$ flips $S^z$, the Kramers degeneracy does not appear within a fixed $(N_\uparrow, N_\downarrow)$ sector. Time-reversal symmetry together with full spin rotational symmetry fixes the Dyson class, namely the Gaussian Orthogonal Ensemble (GOE), once all unitary symmetries are resolved~\cite{Mehta_2004__RandomMatrices}.

\paragraph{Symmetry resolution in exact diagonalization}
We diagonalize $\ham_{\lambda_2,\lambda_3}$ on a finite cluster comprising of $N_x\times N_y$ unit cells in a sector with quantum numbers $(N_\uparrow,N_\downarrow,\vb k, [f])$. The spin-flip quantum number is resolved only for $N_\uparrow = N_\downarrow$. We do not impose fixed total spin $S$ or pseudospin $B$ at the basis level in the numerics; instead, we resolve them using a penalty-term
approach \cite{Poilblanc_1993_EPL_PoissonvsGOE, Teeriaho_2025_Phys.Rev.Res._Coexistenceergodicnonergodic},
diagonalizing $\ham_{\lambda_2,\lambda_3}+\alpha\,\hat{S}^{\,2}+\beta\,\hat{B}^{\,2}$
with sufficiently large positive $\alpha,\beta$ and subsequently selecting the
desired $(S, B)$ subsector, usually the one with the largest dimension. All computations~\cite{Swaminathan_2026__Paritydependentdoubledegeneracya} were performed using the package QuSpin
\cite{Weinberg_2017_SciPostPhys._QuSpinPythonpackage,
Weinberg_2019_SciPostPhys._QuSpinPythonpackage}.

\section{Details on the spectral statistics analysis}
\label{sec:spectral_stats_app}

To quantify spectral correlations of the fully symmetry-resolved many-body spectrum of $\ham_{\lambda_2,\lambda_3}$, we analyze $k$-order level spacings and nonoverlapping  gap ratios. These are useful in the presence of exact degeneracies and multi-block structure. Our analysis follows and extends Refs.~\cite{Oganesyan_2007_Phys.Rev.B_Localizationinteractingfermions, Atas_2013_Phys.Rev.Lett._DistributionRatioConsecutive, Tekur_2018_Phys.Rev.B_Higherorderspacingratios, Rao_2020_Phys.Rev.B_Higherorderlevelspacings, Shir_2025_J.Phys.A:Math.Theor._Surmiserandommatrices}.

\subsection{Ordered spectra and $k$-order spacings}
\label{sec:knn_spacings_app}

For a fully symmetry-resolved sector characterized by quantum numbers $(N_\uparrow, N_\downarrow,\vb k, [f], S, B)$, the ordered eigenvalues $E_1 \le E_2 \le \cdots \le E_{\mathcal N}$ are obtained from exact diagonalization. To reduce the impact of fluctuations near the edges of the spectrum, we discard a small fraction of levels at each end and retain only the bulk spectrum for the level-spacing and gap-ratio analysis. In practice, we discard the lowest and highest fraction $f_{\rm edge} = 0.025$ of eigenvalues.

We define the raw $k$NN spacings by $\tilde{s}_n^{\,k}=E_{n+k}-E_n$ for $k=1,2,3,4,$ computed for all $n$ such that both $E_n$ and $E_{n+k}$ lie in the retained bulk window. Because the density of states varies smoothly with energy, we remove this variation by unfolding, following the procedure of Refs.~\cite{Santos_2004_J.Phys.A:Math.Gen._IntegrabilitydisorderedHeisenberg, Teeriaho_2025_Phys.Rev.Res._Coexistenceergodicnonergodic}. We partition the retained spectrum into $q =30$ disjoint, contiguous sets\footnote{Varying $q$ in the range $[20,100]$ did not affect the inferred statistics.} with an equal number of levels. Within each set, we compute the local mean spacing $\langle \tilde{s}^{\,k}\rangle_{\rm local}$ at fixed $k$. The normalized $k$NN spacings are then
\begin{equation}
s_n^{\,k}=\frac{\tilde{s}_n^{\,k}}{\langle \tilde{s}^{\,k}\rangle_{\rm local}},
\qquad \langle s^{\,k}\rangle_{\rm local}\simeq 1.
\end{equation}
This unfolding procedure is performed independently for each $k$. The corresponding probability density is denoted $P^k(s)$. 

\subsection{Nonoverlapping $k$-order gap ratios}
\label{sec:gap_ratios_app}

As a complementary diagnostic that does not require unfolding, we compute the
nonoverlapping $k$-order gap ratios

\begin{equation}
    r_n^k =
    \frac{\min\bigl(\tilde{s}_n^k,\,\tilde{s}_{n+k}^k\bigr)}
         {\max\bigl(\tilde{s}_n^k,\,\tilde{s}_{n+k}^k\bigr)}
        = \frac{\min\bigl(E_{n+k}-E_n,\,
                    E_{n+2k}-E_{n+k}\bigr)}
         {\max\bigl(E_{n+k}-E_n,\,
                    E_{n+2k}-E_{n+k}\bigr)},
\label{eq:rk_app}
\end{equation}
with $r_n^{\,k}\in[0,1]$ and denote their distribution by $P^k(r)$. 

In the sector with an odd total number of particles $N = N_\uparrow + N_\downarrow$, the spectrum contains exact doublets; consequently, the $k=1$ spacing distribution includes a sharp accumulation at $s=0$ and the nearest-neighbor ratio $r_n^{\,1}$ becomes identically zero for all $n$ (and is therefore not informative). For this reason, and because we are interested in correlations between distinct doublets, we emphasize $k\ge2$ spacing and ratio statistics, which probe correlations between distinct doublets rather than the trivial intra-doublet spacings for odd parity. 

In the presence of exact degeneracies, we also compute a ``collapsed'' version of the spacing data in which exact degeneracies are removed before forming the level spacings,\footnote{In our implementation, eigenvalues closer than a tolerance factor of ${\tt frac}=10^{-11}$ are treated as degenerate for the LSS calculation.} leaving a set of nondegenerate eigenvalues. Unless explicitly stated otherwise, all reported spectral statistics and comparisons use the standard, noncollapsed spacings.

\subsection{Reference RMT distributions and effective $(k^*,\beta^*)$}
\label{sec:rmt_effective_app}

To compare with random-matrix theory (RMT) benchmarks, we use the generalized
Wigner surmise for the $k$NN spacing distribution,
\begin{equation}
P^k(s,\beta)\approx C_{\alpha}\, s^{\alpha}
\exp\!\left[-A_{\alpha} s^2\right],
\label{eq:wigner_k_app}
\end{equation}
where $\beta=1,2,4$ is the Dyson index corresponding to GOE, GUE, and GSE,
respectively. In this approximation, the exponent $\alpha$ is related to $(\beta,k)$ by~\cite{Tekur_2018_Phys.Rev.B_Higherorderspacingratios,
Rao_2020_Phys.Rev.B_Higherorderlevelspacings}
\begin{equation}
\alpha(\beta,k)=\frac{k(k+1)}{2}\beta + k - 1.
\label{eq:alpha_bk_app}
\end{equation}
The coefficients $A_{\alpha}$ and $C_{\alpha}$ are fixed by
normalization and unit mean spacing:
$\int_0^\infty P^k(s)\,ds=1$ and $\int_0^\infty s P^k(s)\,ds=1$, and thus equal to

\begin{equation}
A_\alpha=
\left[
\frac{\Gamma\!\left(\frac{\alpha}{2}+1\right)}
{\Gamma\!\left(\frac{\alpha+1}{2}\right)}
\right]^2,
\qquad
C_\alpha=
\frac{2\,\Gamma^{\alpha+1}\!\left(\frac{\alpha}{2}+1\right)}
{\Gamma^{\alpha+2}\!\left(\frac{\alpha+1}{2}\right)}.
\end{equation}

The Wigner surmise~\eqref{eq:wigner_k_app} with the integer exponent $\alpha$ given by~\eqref{eq:alpha_bk_app} does not reproduce the variance $\Delta_\beta^{(k)}$ of the level spacing distribution $P^k(s, \beta)$ which is known analytically and reads
\begin{equation}
\label{eq:exact_variance}
\Delta_\beta^{(k)} = \frac{1}{k^2}\left(\frac{2 \ln k}{\pi^2\beta} + c_\beta \right)\,.
\end{equation}
The constant $c_\beta$ is the boundary condition, namely the variance for $k=1$. The factor $k^{-2}$ in the above equation appears due to our normalization $\langle s^k \rangle = 1$, which is different from the one in Ref.~\cite{Shir_2025_J.Phys.A:Math.Theor._Surmiserandommatrices}, namely $\langle s^k \rangle = k$. In Ref.~\cite{Shir_2025_J.Phys.A:Math.Theor._Surmiserandommatrices} it is proposed to use for $c_\beta$ the (inexact) variance given of the Wigner surmise~\eqref{eq:wigner_k_app}-\eqref{eq:alpha_bk_app}, namely $c_1 = 4/\pi -1$, $c_2 = 3\pi /8 -1$, $c_4 = 45\pi /128-1$, and to choose the exponent $\alpha$ for $k \geq 2$ by imposing that the variance is given exactly by~\eqref{eq:exact_variance}. This gives a noninteger exponent denoted by $\tilde{\alpha}(\beta, k)$, which is used in place of~\eqref{eq:alpha_bk_app}.
It is found that this \textit{new surmise} leads to a better approximation of the exact $k$NN level spacing distribution compared to~\eqref{eq:wigner_k_app}-\eqref{eq:alpha_bk_app}.

The unfolded $k$NN level spacings computed numerically are fitted to Eq.~\eqref{eq:wigner_k_app}, 
with $\alpha$ as the only free parameter, yielding $\alpha_{\rm fit}$. The effective ensemble Dyson index $\beta^*$ and effective order $k^*$ that best describe the observed distribution are chosen by minimizing $|\alpha_{\rm fit}-\tilde{\alpha}(\beta,k)|$ over $\beta\in\{1,2,4\}$ and $k\le 4$.


For gap ratios, we compare directly to RMT predictions without
unfolding, using known analytical forms for the gap-ratio statistics at $k=1$~\cite{Atas_2013_Phys.Rev.Lett._DistributionRatioConsecutive,
Atas_2013_J.Phys.A:Math.Theor._Jointprobabilitydensities} and their higher-order
generalizations~\cite{Tekur_2018_Phys.Rev.B_Higherorderspacingratios, Rao_2020_Phys.Rev.B_Higherorderlevelspacings}. 

\subsection{Multi-block spectra and $m$-block reference distributions}
\label{sec:multiblock_app}

A common complication in spectral statistics analysis is the presence of multiple
independent blocks that can arise from unresolved symmetries or Hilbert space fragmentation. If the spectrum is a superposition of $m$ statistically independent blocks, levels from different blocks do not repel, and nearest-neighbor statistics can appear closer to
a Poisson-like distribution, even when each block individually exhibits Wigner-Dyson
correlations~\cite{Bleher_2001__Randommatrixmodels,
Anderson_2009__IntroductionRandomMatrices,
Deift_2014__Randommatrixtheory,
Giraud_2022_Phys.Rev.X_ProbingSymmetriesQuantum}. Higher-order diagnostics are particularly useful for resolving such hidden block
structures~\cite{Tekur_2018_Phys.Rev.B_Higherorderspacingratios,
Tekur_2020_Phys.Rev.Res._Symmetrydeductionspectral, Bhosale_2021_Phys.Rev.B_Superpositionhigherorderspacing}.

Accordingly, in addition to single-block comparisons, we compare our data to
multi-block reference distributions $P^k(\xi,\beta,m)$ obtained by
superposing $m$ independent RMT spectra drawn from the $\beta$-Gaussian
ensemble. Here, $\xi=s$ corresponds to the $k$NN spacing distribution, while $\xi=r$ corresponds to the non-overlapping $k$th-order gap-ratio distribution. Throughout, $m$ denotes the number of blocks, and we assume equal
spectral densities for each block in the reference model, i.e., each block
contributes an equal number of levels to the merged spectrum.

\subsubsection{Monte Carlo generation of $P^k(\xi;\beta,m)$}
\label{sec:mc_multiblock_app}

Closed-form expressions for $P^k(\xi,\beta,m)$ are generally not available for
arbitrary $(k,\beta,m)$. We therefore generate reference distributions by Monte
Carlo sampling.

For each $\beta\in\{1,2\}$, we generate independent RMT spectra using the
Dumitriu-Edelman tridiagonal construction~\cite{Dumitriu_2002_J.Math.Phys._Matrixmodelsbeta}. Specifically, we sample $m$ independent $\beta$-ensemble matrices of dimension $N_{\rm RMT} = 1000$ and compute their ordered eigenvalues. We then merge the $m$ eigenvalue lists into a single ordered spectrum and compute $k$NN spacings and ratios from the merged list
exactly as for the many-body data, including the unfolding procedure followed for $P^k(s)$. This procedure is performed $N_{\rm samp} = 1000$ times, and the $P^k(\xi,\beta,m)$ distributions are computed for the combined data~\cite{Swaminathan_2026__Paritydependentdoubledegeneracya}.

\subsubsection{Choosing the closest $(\beta,m)$ reference}
\label{sec:distance_measures_app}

To determine the best matching $(\beta,m)$ for a given set of eigenvalues obtained from exact diagonalization, we compare the
empirical distributions to the Monte Carlo reference distributions and select
the closest match. We use two complementary metrics:

\begin{enumerate}
\item The Kolmogorov-Smirnov (KS) distance between empirical and reference cumulative distribution functions (CDFs)
\cite{Massey_1951_JournaloftheAmericanStatisticalAssociation_KolmogorovSmirnovTestGoodness,
Stephens_1974_JournaloftheAmericanStatisticalAssociation_EDFStatisticsGoodness}.
\item A mean-square deviation (MSD) between empirical and reference
distributions, evaluated on the histogram bins.
\end{enumerate}

Let $\{x_{(i)}\}_{i=1}^N$ denote the sorted data values (either spacings or ratios), and define the empirical CDF evaluated at these points as
\begin{equation}
F_{\mathrm{emp}}(x_{(i)})=\frac{i}{N}\,.
\end{equation}
For an analytic reference curve given as a probability distribution sampled on a grid, we construct the corresponding model CDF by trapezoidal integration and normalization, and then evaluate it at the data points by linear interpolation, yielding $F_{\mathrm{model}}(x_{(i)})$. For Monte Carlo ($m$-block) references, we instead evaluate the reference CDF at the data points via
\begin{equation}
F_{\mathrm{ref}}(x_{(i)})=\frac{1}{N_{\mathrm{ref}}}\,\#\{x^{\mathrm{ref}}_j \le x_{(i)}\}\,,
\end{equation}
implemented by counting the number of reference samples that lie below each data point. In both cases, the Kolmogorov--Smirnov distance is then
\begin{equation}
D_{\mathrm{KS}}=\max_{1\le i\le N}\left|F_{\mathrm{emp}}(x_{(i)})-F_{\mathrm{ref/model}}(x_{(i)})\right|.
\end{equation}

For MSD, we work at the level of histogram densities. Let $\{b_j\}_{j=0}^{B}$ be the bin edges used for the empirical histogram (we use $B=30$ bins over the relevant support interval). Writing $h_j$ for the empirical density in bin $j$, and $m_j$ for the corresponding model/reference density, we define
\begin{equation}
\mathrm{MSD}=\frac{1}{B}\sum_{j=1}^{B}(h_j-m_j)^2.
\end{equation}
For analytic (continuous) reference curves, $m_j$ is obtained by numerically integrating the model probability distribution to form a normalized CDF on a fine grid, interpolating this CDF to the bin edges, and converting the resulting bin probabilities to densities by division by the bin widths. For Monte Carlo $m$-block references, $m_j$ is obtained directly from a histogram of the reference samples evaluated on the same bin edges.

To establish a matching $(\beta, m)$ pair, we require consistency across multiple diagnostics: agreement between $P^k(s)$ and $P^k(r)$, and stability across $k$ values. In ambiguous cases, the gap-ratio statistics are preferred to the level-spacing statistics, as they do not require an unfolding procedure. The KS distance is the primary criterion for selecting the most likely number of blocks $m$, whereas MSD serves as a secondary tiebreaker. 
The effective ensemble index and order $(\beta^*, k^*)$ are inferred from the level-spacing distribution via $\alpha_{\rm fit}$, and the same values are also used in the plots of the nonoverlapping gap ratio. The resulting labels $(k^*,\beta^*)$ or $(k; \beta, m)$ are reported in all the figures and in Tab.~I in the main text.

\section{Supporting Results}

\subsection{Level spacing statistics of $\ham_{0,0}$}

\begin{figure}[h]
  \centering
  \includegraphics[width=0.48\columnwidth]{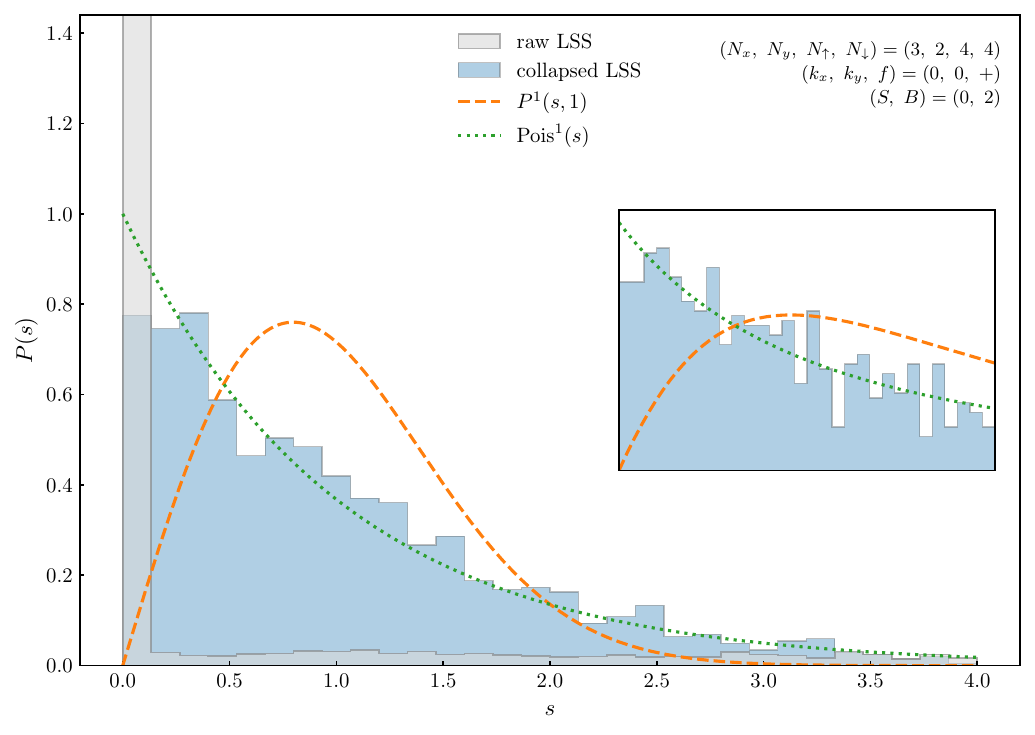}
  \includegraphics[width=0.48\columnwidth]{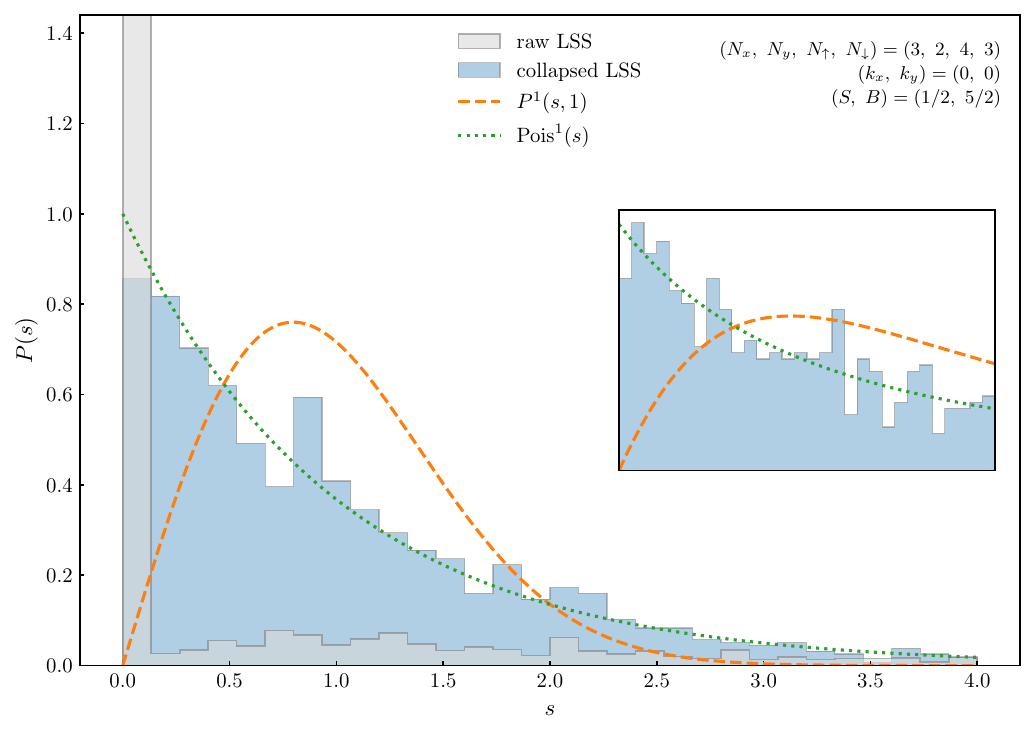}
  \caption{Level spacing statistics for $\ham_{0,0}$ computed on both the raw level spacings and when all degeneracies are collapsed in the even-$N$ sector (left) and odd-$N$ sector (right). The raw level statistics exhibit a so-called ``stick'' structure, characterized by a dominant peak at $s=0$. The collapsed spectrum exhibits level-spacing statistics close to those of the Poisson distribution. The inset shows the gap-ratio distribution for the collapsed spectrum, which also highlights near-Poisson behavior.}
  \label{fig:collapsed_evs}
\end{figure}

As shown in Fig.~1(d) in the main text, the spectrum of $\ham_{0,0} = \ham_{\rm tr.}$ is characterized by highly degenerate multiplets, which are the result of the presence of an extensive number of LIOMs, namely the operators $\hat{S}^2_{n\vb{l}}$ discussed in Sec.~\ref{sec:symmetry_app}. These degeneracies manifest in the level spacing distribution as a large peak at $s=0$, as shown in Fig.~\ref{fig:collapsed_evs}. The level-spacing distribution of $\ham_{0,0}$ does not depend qualitatively on the particle number parity. Similar level-spacing distributions highly concentrated around zero are sometimes called ``stick statistics'' and are encountered in integrable systems built from commuting local projectors~\cite{Kapustin_2020_Commun.Math.Phys._LocalCommutingProjector} and non-interacting systems, for instance spin chains that can be mapped to free fermions by the Jordan-Wigner transformation and collections of harmonic oscillators with commensurate frequencies~\cite{Berry_1977_Proc.A_Levelclusteringregular, He_2026_J.Stat.Mech._Statisticalsignaturesintegrable, Chakrabarti_2012_Phys.Rev.A_Energylevelstatisticsinteracting, Kerin_2025__Quarticlevelrepulsion}.

To probe correlations between distinct many-body energies, we also show in Fig.~\ref{fig:collapsed_evs} the level spacing distribution of the collapsed spectrum introduced in Sec.~\ref{sec:gap_ratios_app}, namely the sequence of energy values obtained by replacing each degenerate multiplet with a single energy level. For
$(\lambda_2,\lambda_3)=(0,0)$, the resulting 1NN spacing and gap-ratio distributions are close to the Poisson benchmark, indicating that once the degeneracies are factored out, the remaining levels are likely
uncorrelated~\cite{Haake_2010__QuantumSignaturesChaos, Guhr_1998_PhysicsReports_Randommatrixtheoriesquantum}.

\subsection{Dependence of average gap ratio on $\lambda_2$ and $\lambda_3$}

\begin{figure}[h]
  \centering
  \includegraphics[width=0.48\columnwidth]{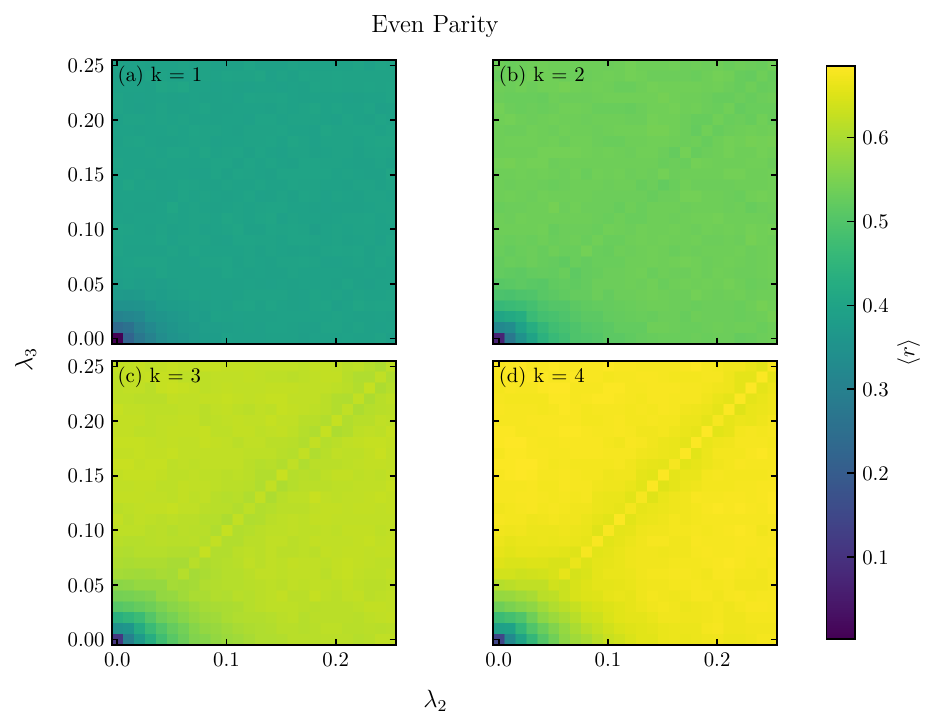}\hfill
  \includegraphics[width=0.48\columnwidth]{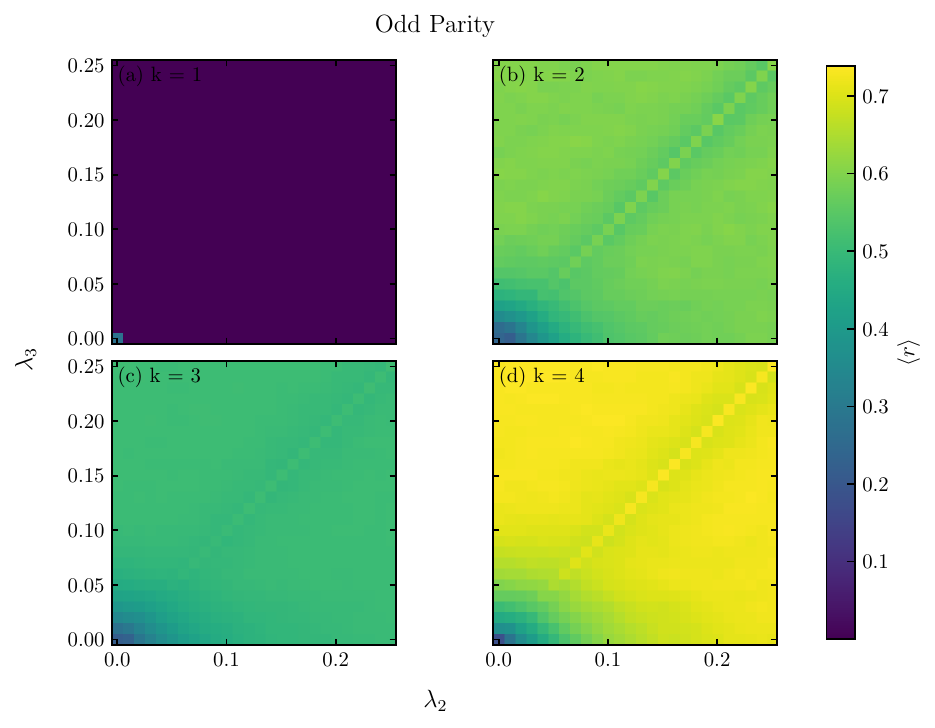}

  \caption{%
    (left): Mean $k$NN  nonoverlapping gap ratio $\langle r^{k}\rangle$ of $\ham_{\lambda_2, \lambda_3}$ as a function of $(\lambda_2,\lambda_3)$ for the even-$N$ sector for lattice size $N_x = 3, N_y = 2$, with $k=1,2,3,4$ in panels (a), (b), (c) and (d), respectively. The point $(\lambda_2,\lambda_3)=(0,0)$ corresponds to $\ham_{\rm tri.}$, where the spectrum is highly degenerate and all $\langle r^{k}\rangle$ are strongly suppressed. Increasing $\lambda_2$ and/or $\lambda_3$ quickly lifts these degeneracies, and the mean gap ratio increases for any order $k$. \,
    (right): Same as (left) but for the odd-$N$ sector. The 1NN ratio $\langle r^{1}\rangle$ in panel~(a) is exactly zero everywhere, reflecting the complete double degeneracy of the spectrum.
    For $k\ge2$ the mean gap ratios increase with $\lambda_2$ and $\lambda_3$, but in contrast to the even-$N$ case, they are nonmonotonic in $k$: away from $(0,0)$ we see that $\langle r^{3}\rangle < \langle r^{2}\rangle,\langle r^{4}\rangle$. Note also the discontinuity of the mean gap ratio along the diagonal. This becomes more visible with increasing order $k$ and is the result of the fact that only for $\lambda_2=\lambda_3$ the pseudospin symmetry is exact and can be resolved in the numerics.  
    }
  \label{fig:heatmap_even_odd}
\end{figure}

Fig.~\ref{fig:heatmap_even_odd} shows the mean nonoverlapping gap ratios $\langle r^{k}\rangle$ across the $(\lambda_2,\lambda_3)$ plane. Near $(\lambda_2,\lambda_3)=(0,0)$, the spectrum is highly degenerate (consistent with the extensive LIOM structure of $\ham_{\rm tri.}$, see Fig.~\ref{fig:collapsed_evs}) and $\langle r^{k}\rangle$ is strongly suppressed regardless of the parity. Turning on $\lambda_{2}$ or $\lambda_{3}$ lifts these degeneracies and leads to the increase of $\langle r^{k}\rangle$. Already for small nonzero $\lambda_i$, the mean nonoverlapping gap ratio displays a strong parity-dependent behavior: in the even-parity sector, $\langle r^k \rangle$ approaches values compatible with the ones expected for uncorrelated eigenvalues or a spectrum composed of several independent blocks. Moreover, the mean gap ratio increases monotonically with the order $k$, $\langle r^k \rangle < \langle r^{k+1} \rangle$. On the other hand, in the odd-parity sector, the 1NN mean gap ratio is identically zero due to the double degeneracy, and one observes the nonmonotonic behavior  $\langle r^{(3)}\rangle <\langle r^{(2)}\rangle,\,\langle r^{(4)}\rangle$.
Finally, both parities show a pronounced feature along $\lambda_2=\lambda_3$, which is the condition for the total pseudospin $B$ to be a good quantum number. Near this line, the total pseudospin symmetry is only approximate, and the associated reduction in level repulsion manifests as a decrease in the gap ratio.  Exactly at $\lambda_2=\lambda_3$, the pseudospin symmetry can be resolved numerically, producing the discontinuity in the mean gap ratio visible in the figure since its value is close to the one attained in the region far from the diagonal.

\subsection{Local observables within degenerate doublets}

\begin{figure}[h]
  \centering
  \includegraphics[width=0.64\columnwidth]{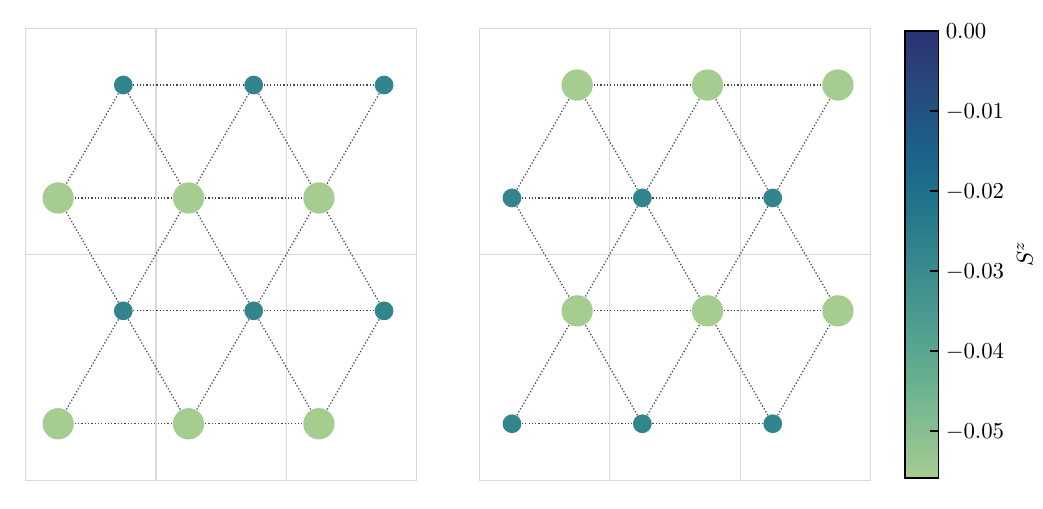}
  \caption{Site-resolved total particle density $\rho_{n\vb{l},i}=\langle \psi_i|\hat\rho_{n\vb{l}}|\psi_i\rangle$ and spin density $S^z_{n\vb{l},i}=\langle \psi_i|\hat S^z_{n\vb{l}}|\psi_i\rangle$ for $\psi_1$ (left) and $\psi_2$ (right) calculated for $(N_x, N_y, N_\uparrow, N_\downarrow, k_x, k_y) = (3,2,2,3,0,0)$ . Here, $\psi_{1,2}$ are two orthonormal states spanning a selected degenerate doublet. The two markers within each rectangular unit cell correspond to the two Wannier orbitals ($n =1,2$). The circle diameter represents the relative particle density at a site, and the spin density is shown on the color scale. The values of $\rho_{n\vb{l}, i}$ and $S^z_{n\vb{l}, i}$ relative to the two states are mapped into each other by a flip in the Wannier index $n=1\leftrightarrow 2$.}
  \label{fig:doublet_density}
\end{figure}

Figure~\ref{fig:doublet_density} shows the site-resolved total particle density
$\rho_{n\vb{l},i}=\langle \psi_i|\hat\rho_{n\vb{l}}|\psi_i\rangle$ and spin
density $S^z_{n\vb{l},i}=\langle \psi_i|\hat S^z_{n\vb{l}}|\psi_i\rangle$ for two
orthonormal states $\ket{\psi_{1,2}}$ spanning a representative degenerate
doublet in the odd-parity sector.
An orbital exchange relates the two states: within each unit cell,
$\rho_{n\vb{l}, i}$ and $S^z_{n\vb{l}, i}$ are mapped into each other by a flip of the Wannier index $n=1\leftrightarrow 2$, for instance $\rho_{1,\vb{l},1}  = \rho_{2,\vb{l},2}$. This suggests
that the states within each doublet are potentially related by a not-yet-identified symmetry that involves the Wannier/band degree of freedom. Choosing a different pair of orthonormal states as a basis for the same degenerate doublet leads to the same result shown in Fig.~\ref{fig:doublet_density} with different specific values of $\rho_{n\vb{l}, i}$ and $S^z_{n\vb{l}, i}$. The same symmetry applies also to any degenerate doublet in the odd-parity sector.

\subsection{Spectral statistics for different system sizes}

\begin{figure}[h]
  \centering
  \includegraphics[width=\textwidth]{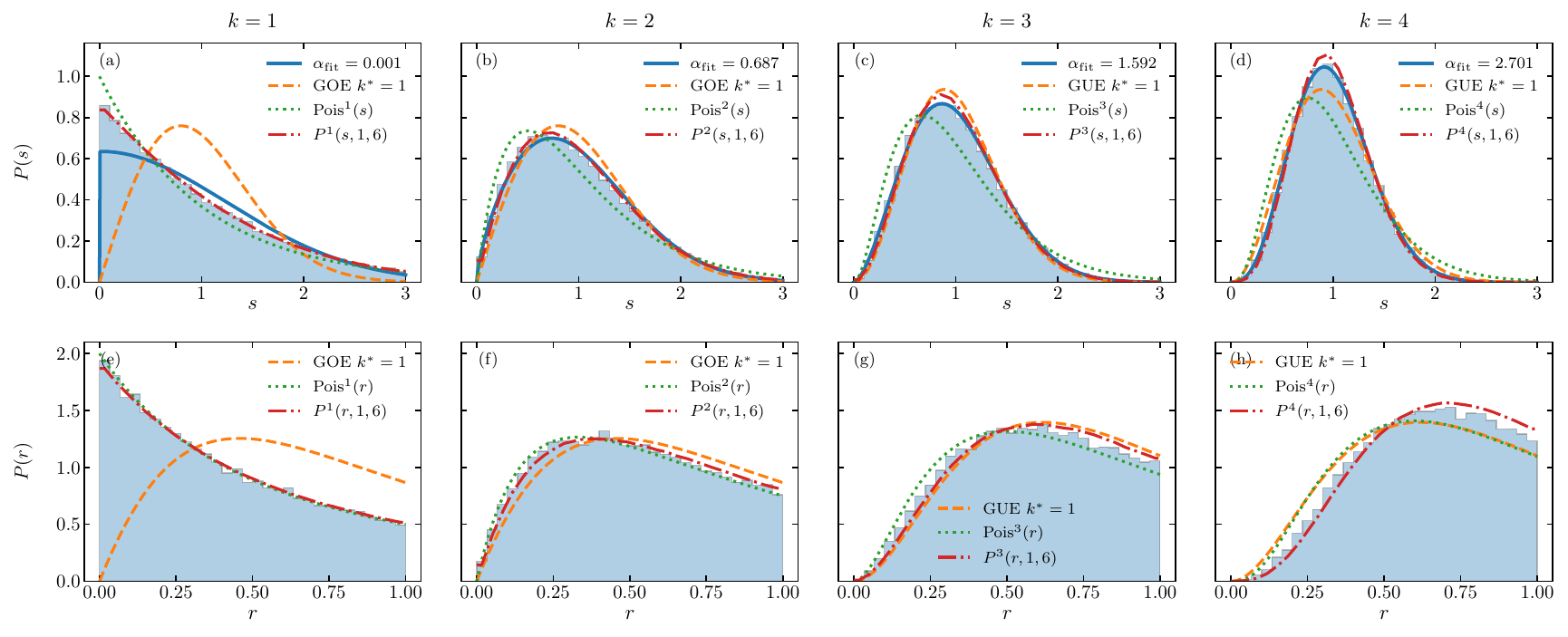}
    \caption{Level spacing (top) and gap ratio (bottom) statistics of $\ham_{1,1}$ for even-$N$, size $N_x, N_y=4, 2$, and quantum numbers $(N_\uparrow, N_\downarrow,k_x,k_y, S, B) = (6,2,0,0,2,5)$ (Hilbert-space dimension $\mathcal N=89652$). In the top row, the solid line shows the fit to \eqref{eq:wigner_k_app}. In both rows, the dashed curve is the closest matching Wigner surmise with effective order $k^*$ (ensemble $\beta^*$). The dotted curve shows the  $k$NN Poisson distribution, and the dash-dot curve is the closest matching $m$-block GOE distribution for each $k$. Both level spacing and gap ratio distributions are consistent with a superposition of $m = 6$ GOE blocks for any $k$.}
  \label{fig:kNNeven_42}
\end{figure}

\begin{figure}[h]
  \centering
  \includegraphics[width=\textwidth]{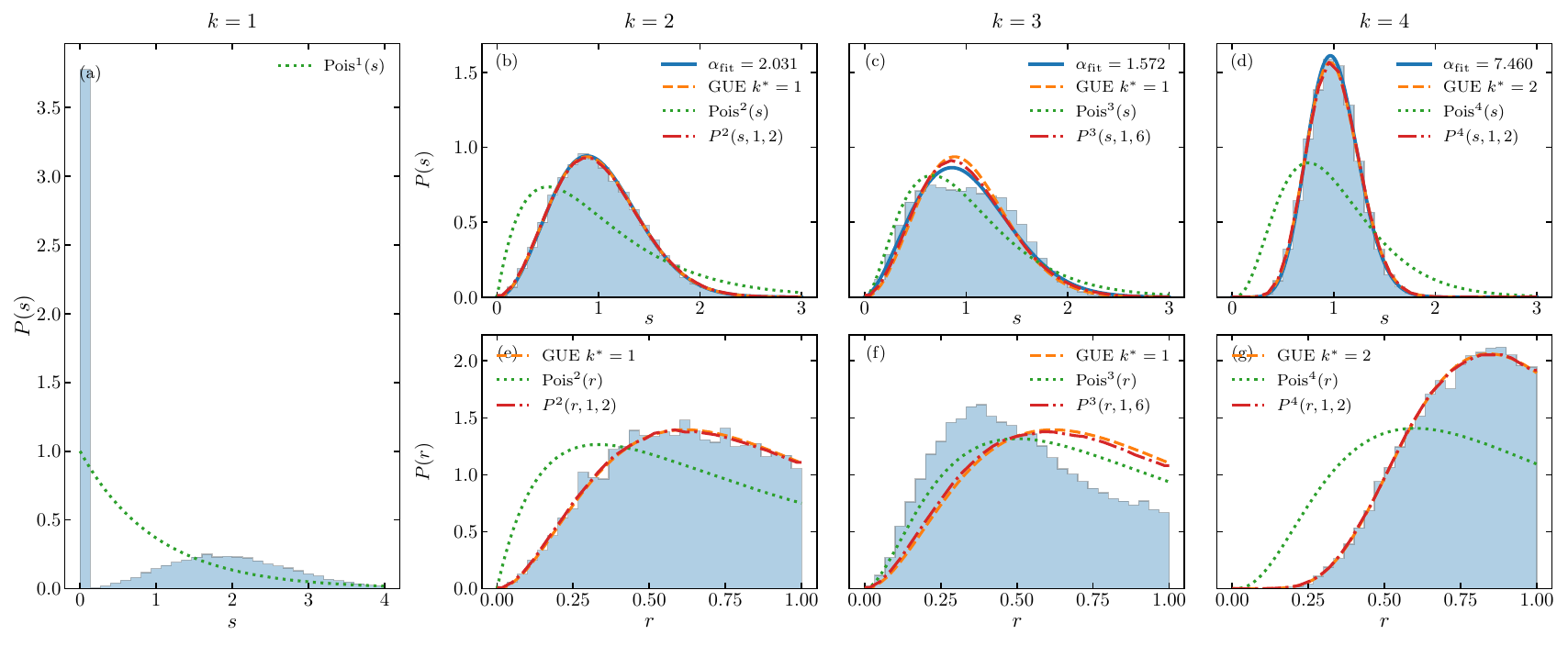}
  \caption{Same as Fig.~\ref{fig:kNNeven_42}, but for odd-$N$ and quantum numbers $(N_\uparrow,N_\downarrow,k_x,k_y,S,B) = (4,3,0,0,1/2,9/2)$ (dimension $\mathcal N=57120$). For $ k=1$, the double degeneracy leads to a pronounced peak at $s=0$ in the level-spacing distribution and identically zero gap ratio~\eqref{eq:rk_app}  (not depicted). The $k=2,4$ statistics are well described by the GUE with $k^* = 1, 2$, equivalent to the superposition of two GOE blocks due to the known relation $P^{2k^*}(\xi, 1, 2)=P^{k^*}(\xi, 2)$. However, the $k = 3$ case clearly deviates from the superposition of two GOE blocks.}
  \label{fig:kNNodd_42}
\end{figure}

\begin{figure}[h]
  \centering
  \includegraphics[width=\textwidth]{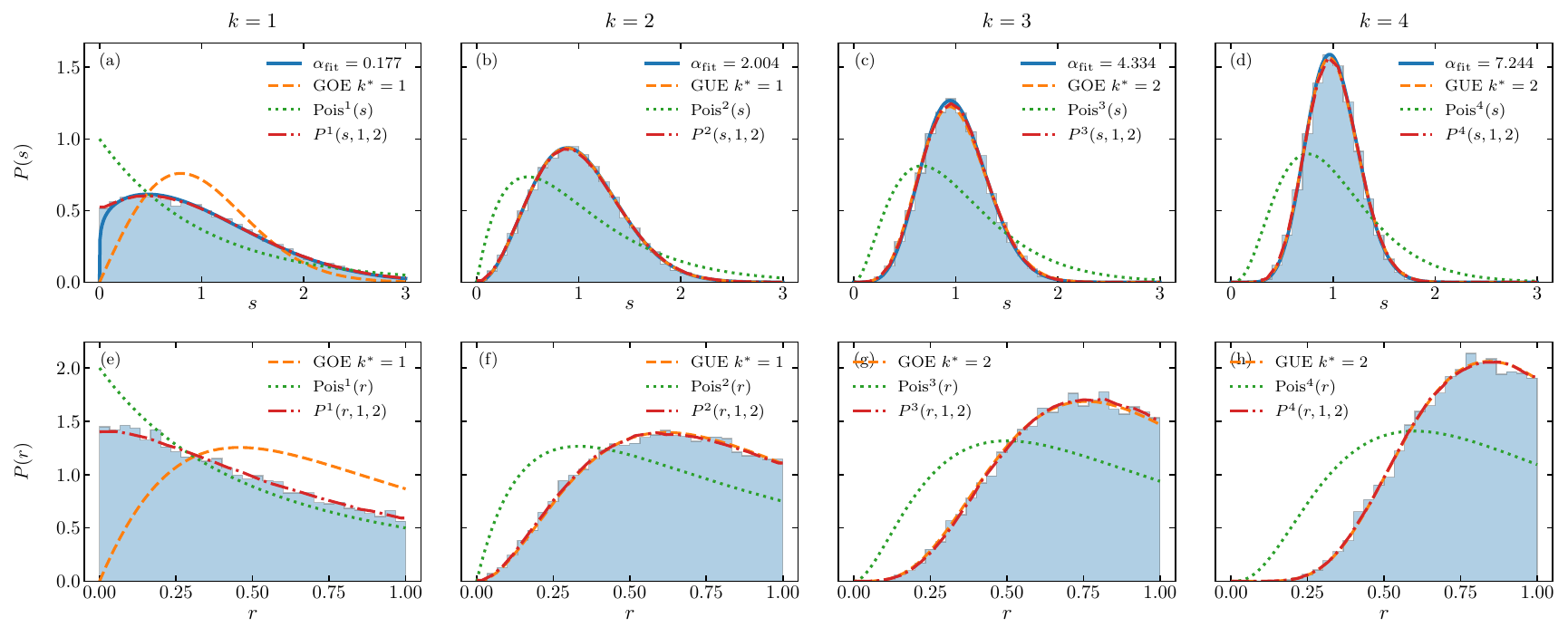}
  \caption{Same as Fig.~\ref{fig:kNNeven_42}, but for size $N_x, N_y= 3, 3$ and quantum numbers $(N_\uparrow,N_\downarrow,k_x,k_y,f,S,B) = (3,3,0,0,+,1,6)$ (dimension $\mathcal N=33592$). Both level-spacing and gap-ratio distributions are consistent with the superposition of two GOE blocks for any $k$.}
  \label{fig:kNNeven_33}
\end{figure}

\begin{figure}[h]
  \centering
  \includegraphics[width=\textwidth]{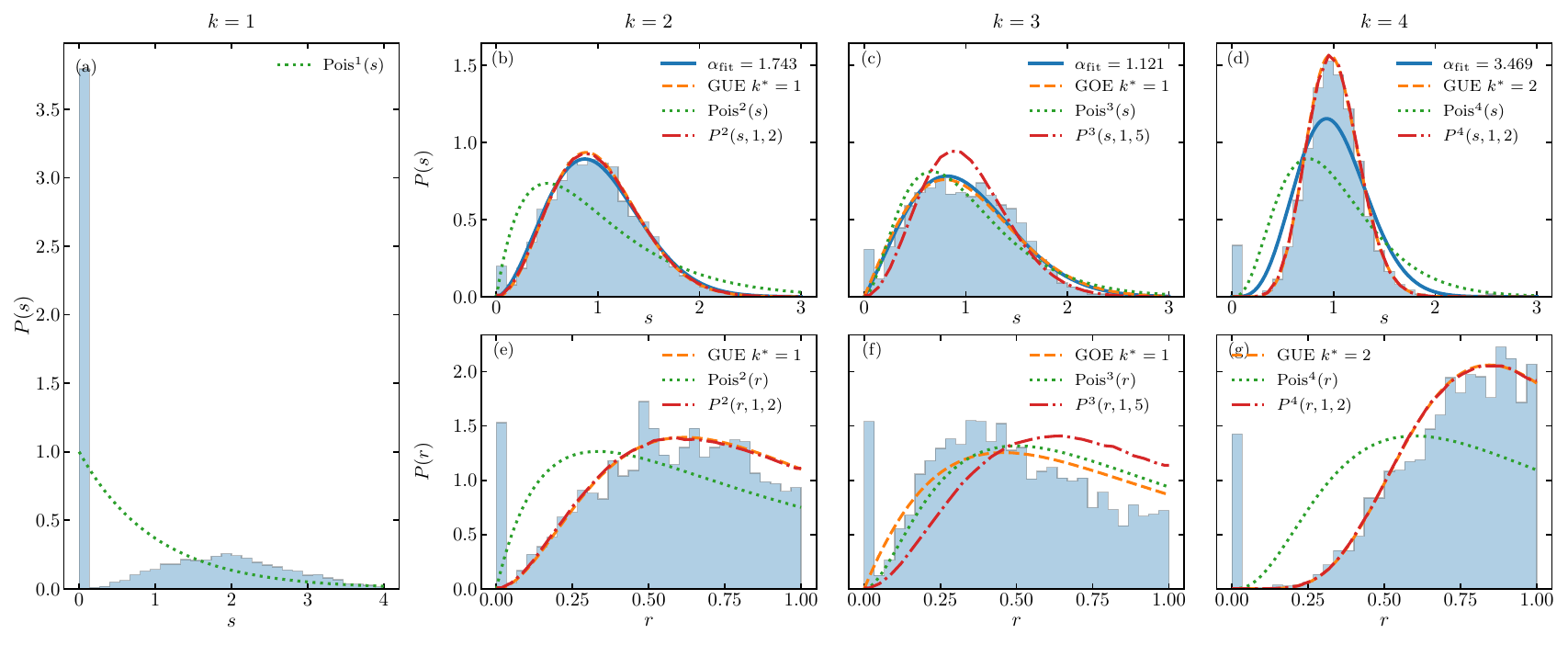}
  \caption{Same as Fig.~\ref{fig:kNNeven_42}, but for size $N_x, N_y= 3, 3$, odd-$N$ and quantum numbers $(N_\uparrow,N_\downarrow,k_x,k_y,S,B) = (3,2,0,0,1/2,13/2)$ (dimension $\mathcal N=7542$). For $k = 1$, the double degeneracy leads to a pronounced peak at $s=0$ in the level spacing distribution and identically zero gap ratio~\eqref{eq:rk_app} (not depicted). The $k=2,4$ statistics are well described by the GUE  with $k^* = 1, 2$, equivalent to the superposition of two GOE blocks (if the large peak at $\xi = 0$ is ignored) due to the known relation $P^{2k^*}(\xi, 1, 2)=P^{k^*}(\xi, 2)$. However, the $k = 3$ case clearly deviates from the superposition of two GOE blocks.}
  \label{fig:kNNodd_33}
\end{figure}

Figs.~\ref{fig:kNNeven_42}--\ref{fig:kNNodd_33} present the spectral statistics analysis for the different system sizes used to fill the entries of Tab.~I in the main text. Fig.~\ref{fig:kNNeven_42} and~\ref{fig:kNNodd_42} show the level-spacing and nonoverlapping-gap-ratio statistics for system sizes $N_x,\,N_y = 4,2$ in the even- and odd-parity sectors, respectively. The results are very similar to the ones for size  $N_x,\, N_y = 3,2$ shown in the main text, with the only significant difference that for even particle number the best matching number of GOE blocks is $m = 6$ for size $N_x,\, N_y = 4,2$ instead of $m=4$. For odd parity, the match with the GUE distribution for $k=2$ and $k=4$ is excellent, as in the case of the data shown in the main text.

The largest system size accessible with exact diagonalization is $N_x,\,N_y = 3,3$. The spectral statistics analysis for this system size is shown in Fig.~\ref{fig:kNNeven_33} (even $N$) and Fig.~\ref{fig:kNNodd_33} (odd $N$). For even parity, the spectral statistics are consistent with a superposition of two GOE blocks. In the odd-$N$ case, the particle number is limited to $N_\uparrow,\, N_\downarrow = 3,2$, for which the dimension of the larger symmetry sector is relatively small, leading to relatively poor statistics compared to the other sizes. Moreover, we observe a pronounced peak at $\xi = 0$, indicating a large number of quasidegenerate states that appear for very low filling. This peak leads to poor fits to the Wigner surmise~\eqref{eq:wigner_k_app}, as shown for instance in Fig.~\ref{fig:kNNodd_33}(d). Thus, only for this figure, the automatic procedure described in Secs.~\ref{sec:rmt_effective_app} and~\ref{sec:multiblock_app} used to identify the best matching numbers $\beta^*$, $k^*$, $m$ is not used for the 2NN and 4NN distributions. Rather, we directly pick the GUE distributions with effective order $k^*=1$ [panels (b) and (e) of Fig.~\ref{fig:kNNodd_33}] and $k^*=2$ [panels (d) and (g)]. These match the numerical data reasonably well if the peak at $\xi = 0$ is ignored, as in the case of the other system sizes reported in Tab. I in the main text. In future work, it will be important to verify that the peak disappears and that the agreement with the GUE distribution improves as the particle number increases while the system size is kept fixed. Achieving this will require more sophisticated numerical methods capable of accessing larger particle numbers, as discussed in the main text.

\subsection*{Effect of a quadratic hopping term}
To test the stability of the observed parity-dependent double degeneracy and spectral statistics, we add to $\ham_{1,1}$ a
single-particle nearest-neighbor hopping term
\begin{equation}
\ham_t=-t\sum_{\langle n\vb{l},n'\vb{l}'\rangle,\sigma}\!\left(\hat{d}^\dagger_{n\vb{l}\sigma}\hat{d}_{n'\vb{l}'\sigma}+\mathrm{H.c.}\right),
\end{equation}
which preserves translational invariance, $\mathrm{SU(2)}$ spin symmetry, spin-flip symmetry, and time-reversal symmetry, and conserves the total particle number, but breaks total pseudospin symmetry. This term lifts the exact double degeneracy for odd parity and removes the multi-block structure observed for even parity. The spectral statistics become essentially parity- and size-independent and are consistent with a single GOE block. This is expected for generic time-reversal and rotationally symmetric Hamiltonians after all symmetries have been resolved. For the results shown in the last column of Tab.~I in the main text, we use $t = 0.1$ for the hopping amplitude.

\clearpage

\bibliography{references}